\documentclass{article}
\usepackage{arxiv}
\usepackage[utf8]{inputenc} 
\usepackage[T1]{fontenc}    
\usepackage{hyperref}       
\usepackage{url}            
\usepackage{booktabs}       
\usepackage{amsfonts}       
\usepackage{nicefrac}       
\usepackage{microtype}      
\usepackage{lipsum}		
\usepackage{graphicx}
\usepackage{doi}
\usepackage{scicite}
\usepackage{longtable}
\usepackage{tcolorbox}

\hypersetup{hidelinks}

\makeatletter
\renewcommand{\section}{%
  \@startsection{section}{1}{\z@}%
                {-2.0ex \@plus -0.5ex \@minus -0.2ex}%
                { 1.5ex \@plus  0.3ex \@minus  0.2ex}%
                {\Large\bfseries\raggedright}
}
\makeatother

\title{LLM-Based Bot Broadens the Range of Arguments in Online Discussions, Even When Transparently Disclosed as AI}

\author{
	Valeria Vuk\thanks{Email: \href{mailto:valeria.vuk@uzh.ch}{valeria.vuk@uzh.ch}.}\\
	Department of Political Science\\
	University of Zurich\\
    \And
    Cristina Sarasua\thanks{Email: \href{mailto:sarasua@ifi.uzh.ch}{sarasua@ifi.uzh.ch}.}\\
	Department of Informatics\\
	University of Zurich\\
	\And
    Fabrizio Gilardi\thanks{Email: \href{mailto:gilardi@ipz.uzh.ch}{gilardi@ipz.uzh.ch}.} \\
	Department of Political Science\\
	University of Zurich\\
}

\renewcommand{\headeright}{}
\renewcommand{\undertitle}{}
\renewcommand{\shorttitle}{LLM-Based Bot Broadens the Range of Arguments in Online Discussions}

\hypersetup{
pdftitle={LLM-Based Bot Broadens the Range of Arguments in Online Discussions, Even When Transparently Disclosed as AI},
pdfauthor={Valeria Vuk, Cristina Sarasua, Fabrizio Gilardi},
pdfkeywords={Online Political Discourse, Large Language Models, LLMs, AI Moderation, Democracy}
}

\begin{document}
\maketitle

\begin{abstract}
A wide range of participation is essential for democracy, as it helps prevent the dominance of extreme views, erosion of legitimacy, and political polarization. However, engagement in online political discussions often features a limited spectrum of views due to high levels of self-selection and the tendency of online platforms to facilitate exchanges primarily among like-minded individuals. This study examines whether an LLM-based bot can widen the scope of perspectives expressed by participants in online discussions through two pre-registered randomized experiments conducted in a chatroom. We evaluate the impact of a bot that actively monitors discussions, identifies missing arguments, and introduces them into the conversation. The results indicate that our bot significantly expands the range of arguments, as measured by both objective and subjective metrics. Furthermore, disclosure of the bot as AI does not significantly alter these effects. These findings suggest that LLM-based moderation tools can positively influence online political discourse.
\end{abstract}


\section*{Introduction}
Broad participation is a cornerstone of democracy, ensuring that many different voices and interests are reflected in political processes. While decision-making processes that allow direct citizen participation often enjoy greater legitimacy \cite{Jacobs.2021, Werner.2022}, their democratic value depends on the consideration of a wide range of perspectives \cite{Goldberg.2021, JASKE.2019}. Digital platforms have the potential to reduce barriers to participation by enabling individuals to engage in discussions anytime and from anywhere. However, disparities in interest, skills, and available time, often linked to socio-economic status, mean that some individuals are less likely than others to engage in political activities on social media, online forums, or e-participation platforms \cite{Hargittai.2016, Oser.2022, Rottinghaus.2020, Baek.2012, Oswald.1.April2025,Schlozman.2018, Griffin.2015, Jacquet.2017, Karpowitz.2014, Karpowitz.2012}. Digital platforms often reinforce homogeneity by encouraging interactions among like-minded individuals \cite{Sunstein.2001, Sunstein.2017} and prioritizing content aligned with existing beliefs \cite{Pariser.2011}. As a result, online discussions may exhibit a limited range of perspectives, which can increase polarization \cite{Yarchi.2021, Wojcieszak.2010,Hobolt:2024a}. These dynamics pose challenges for a healthy democracy.


A limited range of perspectives in online discussions can be addressed through various mechanisms, such as ensuring a representative selection of participants \cite{Griffin.2015, Ryfe.2012}, encouraging less enthusiastic participants to contribute \cite{Kim.2021}, or structuring discussions to ensure balanced participation \cite{Fishkin.2019}. However, no automated approach currently exists for online discussions that accounts for the strong self-selection bias in political discourse. Consequently, recruitment strategies alone are insufficient, and moderators cannot fully compensate for the absence of underrepresented perspectives.

Large Language Models (LLMs) offer new opportunities to improve discourse on digital platforms \cite{.e}. Previous research has shown that integrating conversational agents into online debates can significantly enhance women’s participation \cite{Hadfi.2023}. Integrating AI chat assistants suggesting context-aware rephrasings into online political discussions can significantly enhance perceived conversation quality and democratic reciprocity \cite{Argyle.2023}. When used as a mediator in deliberative settings, AI tools can help groups find common ground \cite{Tessler.2024}.

Building on these studies, we propose ArgumentBot, an LLM-based argument moderator designed to identify and introduce missing arguments into online discussions, thereby broadening the range of perspectives participants are exposed to and potentially engage with. We focus on whether AI should be used in healthcare because it is timely, elicits diverse benefit–risk arguments, and remains open enough for participants to consider unfamiliar perspectives. The bot works with a list of relevant arguments that we compiled. To ensure comprehensiveness, while acknowledging that no list can be entirely exhaustive, we consulted 66 experts from politics, academia, and civil society (e.g., NGOs and civic tech organizations) across Europe. These experts responded to a curated survey with open-ended questions on the use of AI in healthcare. Their responses were independently coded by two co-authors of this paper to identify distinct arguments. The full list of arguments is available in the supplementary material (table S1).

\begin{figure}
    \centering
    \includegraphics[width=0.75\linewidth]{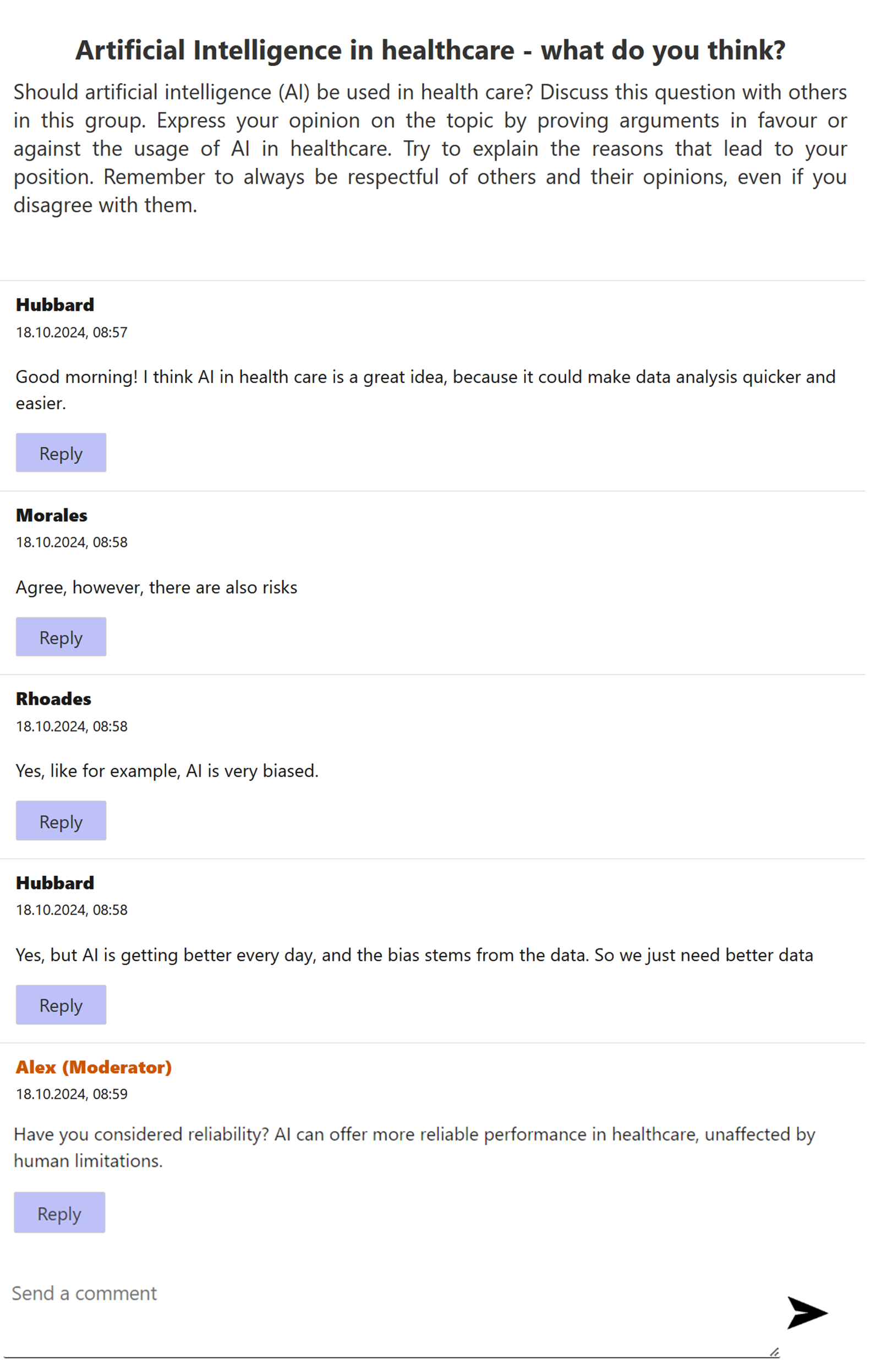}
    \caption{\emph{ArgumentBot at work.} It monitors discussions, identifies missing arguments, and introduces them into the conversation, without explicitly prompting participants to respond.}
    \label{fig:ArgumentBot}
\end{figure}

Figure \ref{fig:ArgumentBot} illustrates how ArgumentBot works in practice. At predefined intervals (2, 5, and 8 minutes) during the discussion, the ongoing conversation is compared to the compiled list of arguments using an LLM. The model identifies missing arguments and randomly selects one to introduce. ArgumentBot then presents the selected argument with the statement “Have you considered [argument]?” followed by a brief explanation prepared by the same two co-authors who coded the survey responses. The exact prompt used for the LLM interaction is provided in the supplementary material (supplementary text). Importantly, ArgumentBot introduces the arguments into the conversation without explicitly prompting participants to respond to them.

To evaluate ArgumentBot’s impact on the discussions, we conducted two pre-registered randomized controlled experiments (N = 1,786 and N = 2,611) in which participants engaged in 10-minute group discussions with three or four other participants, with or without ArgumentBot’s intervention. ArgumentBot was presented in different roles (regular participant, moderator, AI participant, or AI moderator) to assess how its role and labeling influenced discussion dynamics.

We preregistered three key hypotheses, slightly rephrased here without affecting methodology, analyses, or findings. First, we argue that ArgumentBot broadens the range of arguments expressed in the discussion. This does not refer to the overall variety of arguments within a discussion, but more specifically to the unique arguments that participants engage with, either by introducing them or responding to them. 
Mooreover, ArgumentBot may amplify underrepresented perspectives. In group discussions, dominant voices often shape the conversation, while those who perceive themselves as holding minority views tend to remain silent \cite{NoelleNeumann.1974, Asch.1951}. 
While our study does not focus on specific mechanisms, we expect ArgumentBot’s intervention to expand the range of arguments participants engage with. Therefore, our first and main hypothesis is:

\begin{quote}
\textit{H1: ArgumentBot increases the number of unique arguments mentioned by participants in the treatment groups compared to those in the control group.}
\end{quote}

Second, we posit that introducing previously absent arguments can increase the extent to which all participants contribute with similar frequency to the discussion. This claim builds on research on minority opinions, particularly the peer-pressure experiment \cite{Asch.1951} and the spiral of silence theory \cite{NoelleNeumann.1974}. The classic peer-pressure experiment showed that participants often conformed to a unanimous but clearly incorrect majority. However, the presence of even one dissenter significantly reduced conformity \cite{Asch.1951}. The spiral of silence theory builds on the idea people are more likely to express their views when they perceive them as widely shared, and more likely to stay silent when they feel in the minority, which creates a self-reinforcing dominance of one view. Introducing dissent can disrupt this dynamic by legitimizing alternative perspectives and encouraging broader participation \cite{NoelleNeumann.1974}.
Therefore, introducing new arguments could promote more balanced participation:

\begin{quote}
\textit{H2: ArgumentBot leads to more balanced participation among participants in the treatment groups compared to the control group.}
\end{quote}

Third, we contend that introducing missing arguments might make the debate feel more representative of all perspectives. First, a debate that incorporates a wider range of arguments and perspectives is more representative. For individuals who initially hold a minority opinion, seeing their viewpoint acknowledged in the discussion can foster a sense of belonging, which in turn enhances their feeling of being represented. Second, exposure to a wide range of arguments improves participants’ understanding of different views. Recognizing that multiple viewpoints have been considered can further strengthen their sense of being heard and valued. Therefore, our third hypothesis is:

\begin{quote}
\textit{H3: ArgumentBot leads participants in the treatment groups to perceive the discussion as more representative compared to those in the control group.}
\end{quote}

H1 is the hypothesis that directly captures the primary function of ArgumentBot, which is to broaden the range of arguments participants engage with during discussions. The intervention specifically introduces arguments that were previously absent, thereby explicitly targeting the spectrum of perspectives considered. H2 and H3, by contrast, represent secondary, downstream effects that are expected to follow indirectly from this primary intervention. While balanced participation (H2) and perceived representativeness of the discussion (H3) are important outcomes, they rely on the initial broadening of argumentation facilitated by ArgumentBot. However, achieving these downstream effects might involve more complex group dynamics. Therefore, while we expect ArgumentBot to directly broaden argumentation, its overall effect on balanced participation and representativeness is less certain and depends on how these arguments are received and integrated by participants. 

Additionally, we preregistered a research question for each hypothesis to explore how labeling the bot as an AI influences the outcomes. Unlike the hypotheses, these questions do not assume a specific direction of effect. Specifically, we examined whether ArgumentBot’s impact varied depending on whether it was presented to participants as a moderator, a regular participant, or an AI in either of these two roles.


\section*{Results}

\begin{figure}
    \centering
    \includegraphics[width=1\linewidth]{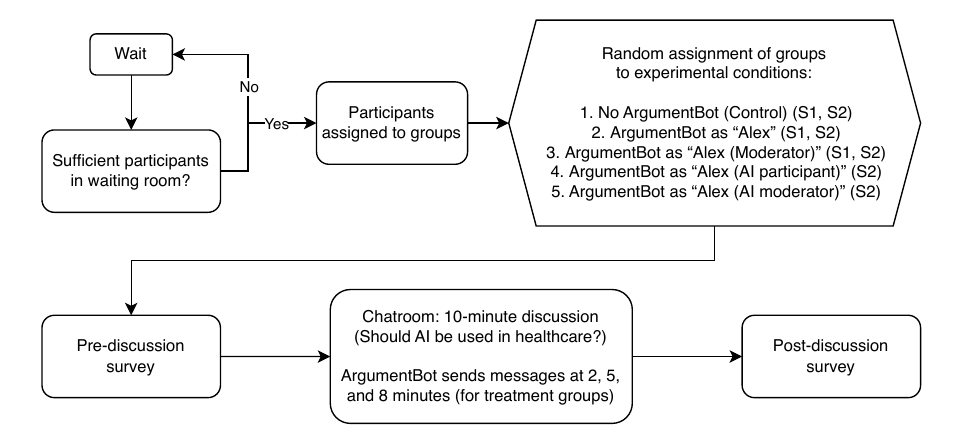}
    \caption{\emph{Overview of the experimental procedure.} Participants were grouped into small discussion groups and randomly assigned to a condition. They completed a pre-discussion survey, discussed the use of AI in healthcare for 10 minutes, with ArgumentBot introducing missing arguments at set intervals in the treatment groups only, followed by a post-discussion survey.}
    \label{fig:experiment}
\end{figure}

We conducted two preregistered between-subjects randomized experiments. Study 1 (N = 1,786) included a control group (no ArgumentBot) and two treatment groups: one where ArgumentBot was labeled ``Alex'' (implicitly, as a participant) and another where it was labeled ``Alex (Moderator)''. Study 2 (N = 2,611) added two additional treatment conditions: ``Alex (AI Moderator)'' and ``Alex (AI Participant)''. Messages from ``Alex (Moderator)'' and ``Alex (AI Moderator)'' were highlighted in orange. Figure \ref{fig:experiment} illustrates the experimental procedure. Participants entered a waiting room for up to five minutes until enough had joined to form groups of four or five participants. Once formed, the groups were randomly assigned to an experimental condition. Before the discussion, they completed a survey on their knowledge of and stance on the discussion topic, attitudes toward AI, political ideology, and socio-demographics. They then spent 10 minutes discussing whether AI should be used in healthcare. In the treatment conditions, ArgumentBot introduced missing arguments at predefined intervals (2, 5, and 8 minutes), while the control group received no bot messages. After the discussion, participants completed a second survey evaluating their experience. For more details on the methodology, see Materials and Methods.

\begin{figure}[t]
    \centering
    \includegraphics[width=1\linewidth]{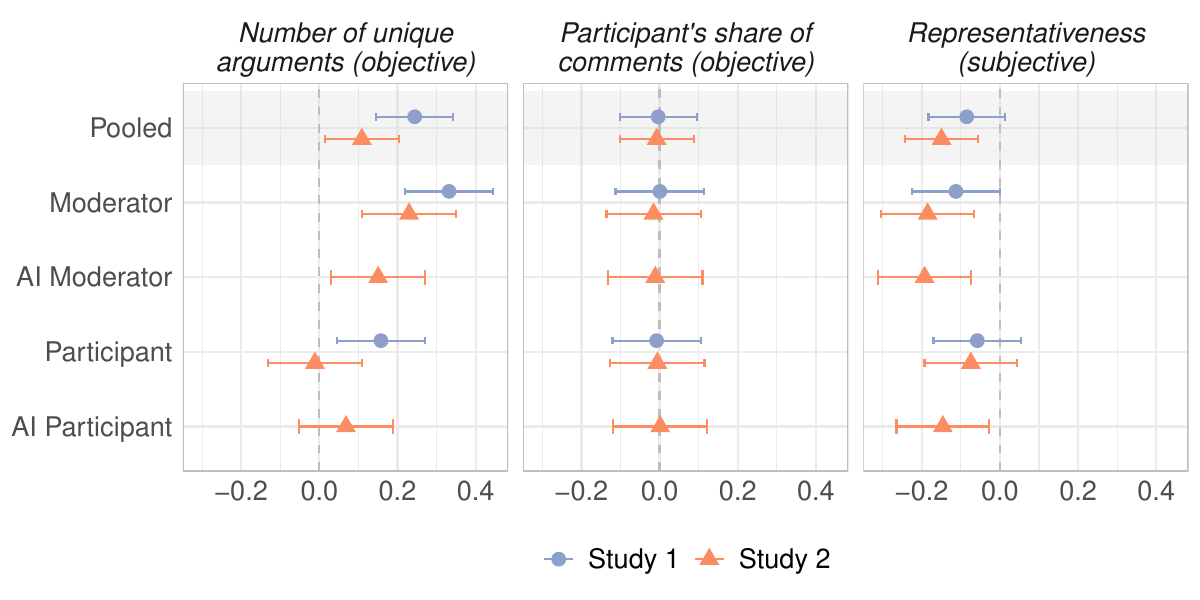}
    \caption{\emph{ArgumentBot increases the number of unique arguments expressed by participants.} The plot shows standardized effect sizes with confidence intervals for our three main outcome measures across Study 1 (N = 1,786) and Study 2 (N = 2,611). The effect sizes are from linear regressions with control variables, with each panel representing a specific measure and separate estimates for different treatment conditions. The horizontal axis shows effect sizes. Pooled estimates summarize the overall effect across all treatment conditions relative to the control.}
    \label{fig:Results}
\end{figure}

We measured our primary outcome (H1) by counting the number of unique arguments each participant mentioned, identified using the predefined argument list. This indicator provides an objective measure of the range of arguments participants express in the discussion. For H2, we measured the outcome by calculating each participant’s comments as a ratio of the group’s average, which is also an objective indicator. By contrast, for H3 we measured the outcome through participants’ perceptions. They rated how well their viewpoints were represented, the extent to which they could express their opinions freely, and whether the discussion represented perspectives from a wide range of groups. As per our pre-analysis plan, the outcome is the mean of these three responses. Figure \ref{fig:Results} presents the standardized effect sizes for these three outcomes across Study 1 and Study 2, comparing the various treatment conditions. The pooled effect represents the overall impact of all treatment conditions combined, relative to the control group. All regression tables can be found in the supplementary material (tables S2-S21).

For H1 and H2, we also measured additional outcomes in the form of subjective assessments collected through a post-discussion survey. Participants rated the range of viewpoints in the discussion and how often they encountered new arguments (H1). They also evaluated the extent to which they felt that individuals from different backgrounds were engaged in the discussion, and whether all participants had a balanced opportunity to contribute (H2). The results for these additional outcomes are shown in figure S1 in the supplementary material.


\subsection*{ArgumentBot increases the number of arguments}

Our main result is that ArgumentBot increases the number of unique arguments expressed by the participants (Figure \ref{fig:Results}), supporting H1. This positive effect was significant in both Study 1 and Study 2 when pooling all treatment conditions (Study 1: 0.244, p = 0.000; Study 2: 0.107, p = 0.026). Specifically, the number of unique arguments exhibited a significant increase when ArgumentBot was presented as a moderator in both studies, although the magnitude was slightly greater in Study 1 (0.332, p = 0.000) compared to Study 2 (0.231, p = 0.0002). In Study 2, ArgumentBot explicitly labeled as an AI moderator also significantly increased argument diversity (0.152, p = 0.013). The role of ArgumentBot as a participant revealed differences between studies: while Study 1 showed a significant positive effect (0.158, p = 0.006), this effect did not replicate in Study 2, where ArgumentBot presented as a regular participant (-0.011, p = 0.859) or an AI participant (0.060, 0.328) did not produce significant effects.

Comparing the number of unique arguments expressed by each group as a whole provides a clearer understanding of the magnitude of these effects. In Study 1, groups in the ``Moderator'' and ``Participant'' conditions both averaged 16.2 distinct arguments (SD = 3.9 and SD = 3.0, respectively). In contrast, the control group averaged only 14.7 distinct arguments (SD = 3.5), reflecting a roughly 10\% increase in the number of arguments. In Study 2, all treatment groups mentioned more distinct arguments than the control. ``AI Moderator'' groups averaged 16.6 distinct arguments (SD = 3.6), while ``Moderator'' groups showed slightly higher diversity at 17.4 (SD = 3.3). Groups where ArgumentBot was presented as a participant had slightly lower values (``AI Participant'': Mean = 16.4, SD = 3.2; ``Participant'': Mean = 16.2, SD = 3.2). The control condition had the lowest number of distinct arguments (Mean = 15.4, SD = 3.0). These results imply that, in Study 2, ``AI Moderator'' groups experienced an 8\% increase in argument diversity, while ``Moderator'' groups saw a nearly 13\% increase compared to the control. Additional details on the distribution of comments and arguments per participant across conditions, as well as the distribution of arguments across groups, are provided in the supplementary material (see figures S2-S7).

In additional analyses, participants were also asked to rate how many new arguments they felt they encountered during the discussion, from their own subjective perspective. Across both studies and in all treatment conditions, participants reported significantly higher exposure to new arguments compared to the control (see supplementary material, figure S1). In Study 1, this perceived increase was significant in both the “Moderator” (0.236, p = 0.000) and “Participant” (0.176, p = 0.002) conditions. Similarly, in Study 2, participants reported greater exposure to new arguments in the “Moderator” (0.164, p = 0.007), “AI Moderator” (0.173, p = 0.004), “Participant” (0.293, p = 0.000), and “AI Participant” (0.189, p = 0.002) conditions. These results also function as a form of manipulation check. Since  ArgumentBot is designed to introduce new arguments into the discussion, it is reassuring to see that participants subjectively noticed and reported this intervention in all treatment conditions.

Finally, we note that our second subjective outcome linked to the number of arguments (perceived openness to a wide range of viewpoints, which is a broader measure than the sheer number of arguments) showed no significant effects in either Study 1 (``Moderator'': 0.023, p = 0.692; ``Participant'': 0.050, p = 0.380) or Study 2 (``Moderator'': -0.016, p = 0.793; ``AI moderator'': 0.037, p = 0.542; ``Participant'': 0.040, p = 0.516; ``AI Participant'': 0.041, p = 0.507) (see supplementary material, figure S1).

\subsection*{ArgumentBot has no effect on participation balance}

Contrary to H2, we found no evidence that ArgumentBot leads to more balanced participation among group members, measured as the ratio of each participant’s comments to the average number of comments within their group. The results from both studies indicate that ArgumentBot had no significant effect on this outcome (Study 1 -- moderator: 0.001, p = 0.990; participant: -0.008, p = 0.894; Study 2 -- moderator: -0.015, p = 0.804; AI moderator: -0.011, p = 0.859; participant: -0.006, p = 0.927; AI participant: 0.002, p = 0.979). This finding remains unchanged when using the number of tokens instead of the number of comments (see supplementary material, table S21). 

Regarding our additional subjective measures, in Study 2, ArgumentBot had a significant negative effect on participants’ perceptions of the opportunity to participate (``Moderator'': -0.203, p = 0.001; ``AI moderator'': -0.163, p = 0.008; ``Participant'': -0.144, p = 0.019; ``AI participant'': -0.121, p = 0.049) (see supplementary material, figure S1). However, no such effect was observed in Study 1 (``Moderator'': -0.021, p = 0.722; ``Participant'': -0.025, p = 0.671). In both studies, ArgumentBot had no significant effect on perceptions that participants from different backgrounds were equally involved (Study 1: ``Moderator'': -0.073, p = 0.201; ``Participant'': -0.051, p = 0.366; Study 2: ``Moderator'': -0.080, p = 0.183; ``AI Moderator'': -0.091, p = 0.132; ``Participant'': -0.041, 0.493; ``AI Participant'': -0.021, p = 0.724).

\subsection*{Negative effect of ArgumentBot on perceived representativeness?}

Against H3, we find mixed evidence that ArgumentBot may decrease, not increase, the perceived representativeness of the discussion (Figure \ref{fig:Results}). In Study 2, ArgumentBot led to significantly lower subjective ratings of representativeness in most conditions  (``Moderator'': -0.185, p = 0.002; ``AI Moderator'': -0.193, p = 0.002; ``Participant'' -0.074, p = 0.222; ``AI participant'' -0.148, p = 0.015). Results from Study 1 are similar (``Moderator'': -0.113, p = 0.050; ``Participant'': -0.058, p = 0.313). However, the subjective perceptions of participants cannot be independently validated, as we have no objective measure for this outcome.

\subsection*{Labeling ArgumentBot as AI does not affect the results}

Both studies show that the AI label did not influence the main outcome (the number of unique arguments expressed by participants). Pairwise comparisons between treatment conditions indicate that the presence or absence of an AI label had minimal influence, with ArgumentBot’s effects remaining largely consistent across conditions. In Study 2, which introduced the AI labels, there were no statistically significant differences between AI-labeled and non-AI conditions (“AI Moderator” vs. “Moderator”: -0.079, p = 0.573; “AI Participant” vs. “Participant”: 0.080, p = 0.568). When comparing roles, moderators had a stronger effect than participants in conditions without the AI label (``Moderator'' – ``Participant'': Study 1: 0.174, p = 0.002; Study 2: 0.241, p = 0.001). In contrast, no such difference was observed in Study 2 when both were labeled as AI (“AI Moderator” – “AI Participant”: 0.082, p = 0.541). The full pairwise comparisons between treatment conditions for our main outcome are reported in table S22 in the Supplementary Material.


\subsection*{Potential mechanisms}

An important finding is that labeling the bot's messages as AI did not significantly alter the effects. This result is particularly interesting because one might expect that knowing a message comes from a bot would influence how participants engage with it. One possible explanation is that the sender's identity may not be as influential as commonly assumed \cite{Broockman.2024}. Participants may not have considered the bot’s ``AI identity'' as significant enough to change their engagement. Another possibility is that as people become more accustomed to engaging with AI in everyday contexts, they may start to treat it as a normal part of the online experience.

Participants in the treatment groups reported encountering a broader range of arguments and also introduced a greater variety of arguments into the discussion, even though the bot did not explicitly prompt participants to respond to its suggestions. This effect may be due to participants directly adopting the arguments provided by the bot or because ArgumentBot’s messages stimulated new ideas and perspectives. The latter dynamic is illustrated in the conversation snippet in the supplementary material (supplementary text, snippet 1).

Despite its success in broadening the range of arguments, ArgumentBot negatively affected perceived balance and representativeness while having no significant impact on actual participation balance. One possible explanation lies in the nature of the discussions themselves. In both studies, participants engaged in brief, 10-minute discussions. Many respondents, as noted in comments from the second survey, felt that the limited time constrained their ability to fully engage. Moreover, ArgumentBot suggested arguments that were not always directly aligned with the ongoing exchange. This effect was particularly pronounced when ArgumentBot was explicitly labeled as an (AI) moderator or AI participant, suggesting that its visible presence may have distracted participants or drawn their attention toward engaging with the bot’s arguments rather than generating their own. 
This difference in engagement is evident in the example snippet in the supplementary material (supplementary text, snippet 2) from a discussion with Alex (implicitly presented as a participant).

Another possible explanation for the negative impact on perceived representativeness relates to the nature of the arguments contributed by ArgumentBot. While ArgumentBot was designed to introduce arguments that had not been mentioned before, it did so without considering how they fit into the broader structure of the discussion. 
Moreover, our list of arguments comprises expert or mainstream ideas but does not include fringe opinions, which may make participants holding those views feel less represented. Finally, a broad range of arguments may not necessarily enhance the quality of the exchange if participants perceive them as external intrusions rather than as meaningful or relevant contributions.

Our analysis indicates that the objectively balanced participation does not align with subjective evaluations. A possible explanation is the short duration of the discussions, which may not have provided sufficient time for participation patterns to shift. Additionally, since participants were incentivized to contribute in order to receive payment, this external motivation may have encouraged more uniform engagement across all conditions. The discrepancy between perceived and actual balance presents an important avenue for future research and has implications for the design of online discussion platforms.

\section*{Discussion}

Our study demonstrates that ArgumentBot, the LLM-based argument moderator we developed, effectively increased the number of unique arguments participants engaged with in an online discussion, as reflected in both objective measures and participants’ subjective perceptions. ArgumentBot was most effective when acting as a moderator rather than as a regular participant, with groups in the treatment conditions expressing approximately 8–13\% more unique arguments compared to groups in the control condition. However, this increase in the number of unique arguments came with potential trade-offs. Despite its success in broadening the range of perspectives, ArgumentBot did not make participation more (nor less) balanced. Moreover, its presence did not enhance, and in some cases even diminished, perceived representativeness and opportunity to contribute. These findings underscore a potential tension in the deployment of AI in social contexts: while AI can enrich online discourse by introducing previously unconsidered arguments, it may simultaneously disturb participants’ sense of efficacy in the conversation.

A key finding is that disclosing ArgumentBot as AI did not significantly alter either participants’ perceptions or objective measures related to the three dimensions we investigate. Whether the bot was presented as human or AI had no measurable impact on the range of arguments, balance of participation, and perceived representativeness. While a recent survey experiment suggests the presence of an AI penalty in online discussions based on hypothetical scenarios \cite{Jungherr.10.03.2025}, our behavioral results of exposure to an actual LLM-based bot find no such negative impact, even when the use of AI is disclosed transparently. This is particularly relevant in the context of legislation mandating transparency in AI usage, such as the European Union's AI Act. Ensuring that users are not misled in online discussions is essential, and our results indicate that transparency about AI did not diminish ArgumentBot’s positive effect on the range of arguments in our experiments. Moreover, the fact that participants’ subjective perceptions remained largely unchanged by labeling bots as AI suggests that such disclosure does not significantly influence the overall user experience. This finding supports the feasibility of integrating AI-driven argument moderators into online discussions without concerns that transparency requirements will undermine their effectiveness.

Our study has several limitations. First, the discussions were brief (10 minutes), which many participants felt constrained their ability to fully express their views. Second, the synchronous, small-group format (4–5 participants) may not reflect the dynamics of large-scale, asynchronous political discussions. Platform-specific norms and features may also shape how ArgumentBot is received. Third, the bot’s intervention strategy could be improved by responding more dynamically and incorporating previous comments. Fourth, because representativeness is subjective, we lacked an objective measure of it. Given the bot’s negative effect on perceived representativeness, this warrants further investigation. Fifth, we focused on a relatively open topic (AI in healthcare) where participants may be more receptive to new arguments. It remains to be seen how ArgumentBot performs in more polarized contexts where views are more entrenched.

Despite these limitations, our findings have important implications. Engagement with a wide spectrum of arguments is essential for a healthy democracy, yet online spaces often restrict the range of perspectives due to self-selection of participants and platform design choices. Our study suggests that ArgumentBot can help counter these dynamics by introducing a broader set of viewpoints, even when transparently disclosed as AI, fostering a more open exchange of ideas, and supporting more balanced participation in public debates. By enriching the variety of arguments in online discussions, ArgumentBot has the potential to contribute to a more robust and democratic discourse.

\section*{Materials and Methods}

\subsection*{Experimental Design}

We conducted two preregistered between-subjects randomized controlled trials (Study 1: July 2024, Study 2: December 2024). The treatment in both experiments was an LLM-based argument moderator that we developed to introduce previously unconsidered arguments during group discussions (ArgumentBot). The design of Study 1 included a control group (no ArgumentBot) and two treatment groups: one in which ArgumentBot was referred to as ``Alex'' (Treatment 1) and another where it was labeled ``Alex (Moderator)'' (Treatment 2). In Study 2, two additional treatment conditions were introduced: ``Alex (AI Moderator)'' (Treatment 3) and ``Alex (AI Participant)'' (Treatment 4). Messages from ``Alex (Moderator)'' and ``Alex (AI Moderator)'' were highlighted in orange.

The experimental procedure is shown in Figure \ref{fig:experiment}. Participants first entered a waiting room, where they remained for a maximum of five minutes or until enough participants had joined. Once a group was formed, it was randomly assigned to one of the experimental conditions with equal probabilities. Randomization occurred at the group level, as participants had to wait for a sufficient number of others to enable a synchronous discussion. Before the discussion, participants completed an initial survey. 
The discussion took place on the platform 'Tool for Online Discussions' \cite{Hoes.2023}, lasted 10 minutes and focused on the question: Should AI be used in healthcare? Most groups consisted of five participants, with a few having four. In the treatment conditions, ArgumentBot introduced arguments that had not yet been mentioned at predefined intervals (2, 5, and 8 minutes). In the control condition, no messages from ArgumentBot were provided. Following the discussion, participants completed a second survey assessing their experience during the conversation.

\subsection*{Data and Sample}

All participants were recruited via Prolific. Recruitment was limited to the UK. Participants were required to be at least 18 years old and fluent in English. We did not aim for a sample balanced across demographic characteristics, as our experiment assumes that online participation is inherently shaped by self-selection. An unbalanced sample, therefore, better reflects the population of interest. Participants were excluded from the analysis if they: (1) failed one of two attention checks; (2) experienced technical issues that prevented them from completing the discussion; or (3) were placed in a group with fewer than four participants. 
The first attention check was administered before the group discussion, while the second was given after the discussion. The final sample included 1,786 participants in Study 1 and 2,611 participants in Study 2. Additional demographic information is available in the supplementary material, table S23.

\subsection*{Outcome measures}

\emph{Number of unique Arguments} (hypothesis 1) was measured by counting the number of unique arguments each participant mentioned.  After removing all bot-generated comments, we used GPT-4o to analyze each participant’s comments and determine whether they contained arguments from our predefined list (the full prompt is provided in the supplementary text). To validate the procedure, we randomly selected 100 comments from both studies and compared its annotations to independent assessments by two of the co-authors.  The two coders agreed with GPT-4o’s annotations in 80\% of the cases and did not identify a specific argument that GPT-4o repeatedly labeled incorrectly. We also collected subjective assessments through a post-discussion survey. Participants were asked to rate the extent to which the discussion allowed for a diverse range of viewpoints (``Range of viewpoints seen'') on a scale from 1 (lack of different viewpoints) to 5 (wide diversity of viewpoints). They were also asked how often they encountered arguments they had not considered before (``New arguments seen''), using a scale from 1 (never) to 5 (very often).

\emph{Participant's share of comments} (hypothesis 2) was measured as the ratio of each participant’s comments to the mean number of comments within their group. This approach allows for a standardized comparison across groups with varying total comment counts \cite{Karpowitz.2014}. The same calculation was also performed at the token level, with the results fundamentally unchanged (see supplementary material, table S21). Additionally, participants were asked to evaluate their perceived balance of participation. The variable ``different backgrounds'' assessed the extent to which they felt that participants from different backgrounds were equally engaged in the discussion (1 = not at all engaged, 5 = extremely engaged). ``Opportunity to participate'' measured perceptions of whether all participants had an equal opportunity to contribute and share their thoughts (1 = not at all, 5 = completely).

\emph{Representativeness} (hypothesis 3) was measured exclusively through participants’ perceptions. They were asked three questions: how well they felt their own viewpoints and opinions were represented in the discussion (1 = not represented at all, 5 = very well represented), the extent to which they felt able to express their opinions freely (1 = not at all, 5 = very well), and whether participants felt the discussion included perspectives from underrepresented or marginalized groups (1 = not at all, 5 = very much). Following our pre-analysis plan, our measure of representativeness is the mean of these three survey responses.

\subsection*{Statistical Analysis}

All dependent variables were standardized using z-scores to facilitate comparison. Each hypothesis was tested using linear regression models with control variables, including group size, age, gender, education, experience in political discussions, and experience in online discussions. Results are robust to using negative binomial and multilevel models. All regression tables are available in the supplementary material (tables S2-S21).

\newpage
\bibliographystyle{sciencemag}
\bibliography{references}

\newpage

\subsection*{Acknowledgments}
We thank Yating Pan for outstanding research assistance, and Colin Henry and Hannah Werner for helpful comments.

\subsection*{Funding}

V.~V. and C.~S. were funded by the Swiss National Science Foundation (SNSF) under grant ID CRSII5\_205975 (D3 Project). F.~G. was funded by the European Research Council (ERC) under the European Union's Horizon 2020 research and innovation program (grant agreement nr. 883121). F.~G. and V.~V. were funded by a generous grant by the Department of Political Science, University of Zurich.

\subsection*{Author contributions}
V.~V. defined the overarching research goals and performed data curation. V.~V., F.~G., and C.~S., designed the research, performed analyses, and wrote the paper.


\subsection*{Competing interests}
There are no competing interests to declare.

\subsection*{Data and materials availability}

Data and code to reproduce all analyses have been deposited in Harvard Dataverse (doi:10.7910/DVN/JVCARG  ) and will be made available upon publication.

\newpage


\renewcommand{\thefigure}{S\arabic{figure}}
\renewcommand{\thetable}{S\arabic{table}}
\renewcommand{\theequation}{S\arabic{equation}}
\renewcommand{\thepage}{S\arabic{page}}
\setcounter{figure}{0}
\setcounter{table}{0}
\setcounter{equation}{0}
\setcounter{page}{1} 


\begin{center}
\section*{Supplementary Materials for: LLM-Based Bot Broadens the Range of Arguments in Online Discussions, Even When Transparently Disclosed as AI}

Valeria~Vuk,
Cristina~Sarasua,
Fabrizio~Gilardi
\end{center}

\subsubsection*{This PDF file includes:}
Supplementary Text\\
Figures S1 to S7\\
Tables S1 to S23\\


\newpage










\subsection*{Supplementary Text}




\subsubsection*{GPT Prompt for ArgumentBot}\label{text:bot}

Here is the log data from \${startTime}:00 to \${endTime}:00:
\${preprocessedLog}

Here is the list of arguments with brief a explanation:
\${argumentsList.join(', ')}

Your tasks are:
\begin{itemize}
    \item Identify and list all the arguments in the list that mentioned in the 
    current log using \textless arguments\_mentioned\textgreater arguments mentioned here \textless /arguments\_mentioned\textgreater, 
    separating arguments with commas. If no arguments are mentioned, use \textless arguments\_mentioned\textgreater  None \textless/arguments\_mentioned\textgreater .
    \item Identify and list all the arguments in the list that are not mentioned from the beginning to the current log using \textless arguments\_not\textgreater arguments not mentioned here \textless /arguments\_not\textgreater , separating arguments with commas.
    \item Pick one of the missing arguments randomly from the list of arguments not mentioned, and insert it into "Have you considered [selected\_missing\_argument]?" with the brief explanation of the argument from the list. Don't add any special characters or punctuation around the argument.
\end{itemize}

\newpage 

\subsubsection*{GPT Prompt for calculating the 'Number of unique arguments' variable} \label{text:diversity_calculation}

SYSTEM\_PROMPT\_TEMPLATE = ''' \\
              You are an ArgumentBot for democratic discussions about AI applications in medicine and health. 
              
              Given the following list of arguments (LoA): 
              \\n \\
              \{\}
              \\n
              
              Your task is to identify the arguments from the list of arguments (LoA) that are explicitly mentioned in a comment that you will be given. The comment was written by a user in a discussion.\\
              You need to return:\\
              - The name of the argument\\
              - An explanation about your classification\\
              Please write the JSON output as {{"arguments": [{{"name": "the name of the argument", "explanation":"the explanation"}}, {{"name": "the name of the argument", "explanation":"the explanation"}}]}}\\
              If you think the comment does not really contain an argument, then write "I cannot find an argument in this piece of text." as the name of the argument and leave the explanation empty.
              You must output a valid JSON.\\
              
              '''\\
 
labels = []\\
    
USER\_PROMPT\_TEMPLATE = ''' \\
                What arguments from the list of arguments (LoA) do you identify in the following comment?\\
                \\n\\
                \{\}
               \\n\\
                '''

\newpage 

\subsubsection*{Snippet 1 from group discussion (group number 423, study 2)} \label{text:snippet1} 
... \\
\textbf{Baldwin:} it would be useful if it was used for example to analyse vital signs chats for example. \\
\textbf{Baldwin:} charts \\
\textbf{Lujan:} I think it would be for imaging and working out medication doses etc \\
\textbf{Comer:} also not sure how empathetic a chatbot would be \\
\textcolor{orange}{\textbf{Alex (Moderator)}}: Have you considered identification of rare symptoms? AI is good at spotting rare symptoms in healthcare data that humans might miss. \\
\textbf{Baldwin:} exactly ai will have much larger databases \\
\textbf{Baldwin:} it can replace a nurse in terms of detecting problems \\
\textbf{Comer:} Do you worry about mass unemployment of nurses then? \\
\textbf{Baldwin:} no because nurses are always needed \\
... \\


\subsubsection*{Snippet 2 from group discussion (group number 86, study 2)} \label{text:snippet2} 
... \\
\textbf{Terry:} I dont feel AI should be used in the healthcare sector as its too unreliable it will also cut down alot of jobs and make the countrys society alot worse when it comes too unemployment 
 \\
\textbf{Woodruff:} good point \\
\textbf{Alex:} Have you considered the healthcare system integration challenge? AI integration into existing healthcare frameworks can be challenging. \\
\textbf{Bravo:} I agree with Morales, it is going to be used regardless so we should work with it and take advantage of it. It can improve efficiency \\
\textbf{Morales:} I've worked in healthcare and there are so many inefficiences and  delays because we rely on humans for everything! Realistically on the ground, there are lots of things that might not need a human hand, although human oversight, yes. \\
\textbf{Terry:} whilst it can help in ways that it can cut down time and help nhs staff i still dont 100\% want too put my trust in a robot  \\
\textbf{Morales:} As for jobs, new ways of working (and therefore jobs) may well be created, it's not just losses.\\
\textbf{Terry:} as in what way? \\
\textbf{Mcelroy:} As someone who works in healthcare I think AI could definitely help in areas of training and admin duties, that would free up time for staff to actually do the job they are there to do. \\
... \\


\newpage
\subsection*{Supplementary Figures \& Tables}

\begin{figure}[ht]
    \centering
    \includegraphics[width=1\linewidth]{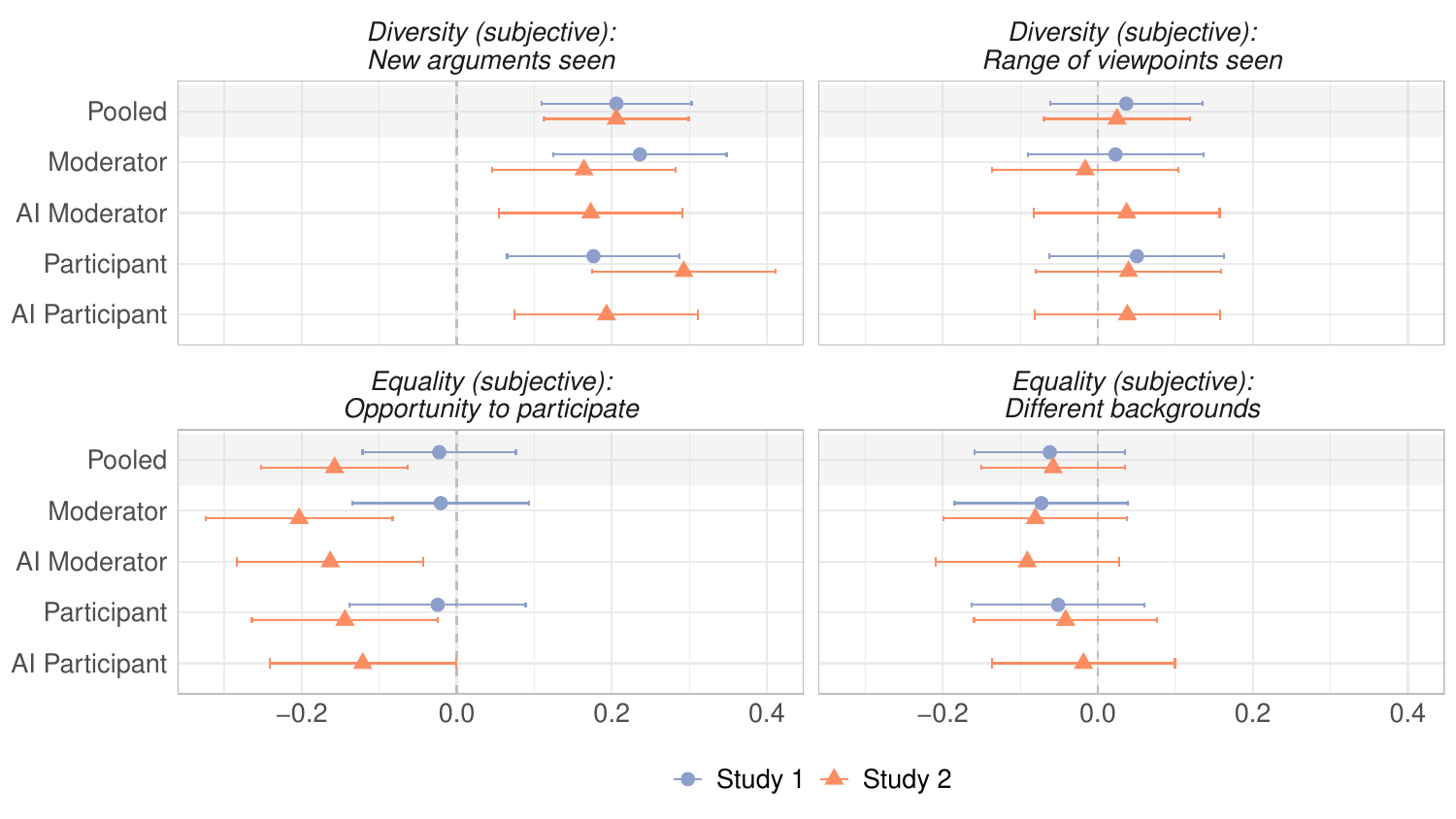}
     \caption{\emph{Effects for additional subjective outcomes.} The plot shows standardized effect sizes with confidence intervals for additional subjective outcome measures across Study 1 (N = 1,786) and Study 2 (N = 2,611). The effect sizes are from linear regressions with control variables, with each panel representing a specific measure and separate estimates for different treatment conditions. The horizontal axis shows effect sizes. Pooled estimates summarize the overall effect across all treatment conditions relative to the control.}
     \label{fig:results_additional}
\end{figure}

\newpage

\begin{figure}[ht]
    \centering
    \includegraphics[width=1\linewidth]{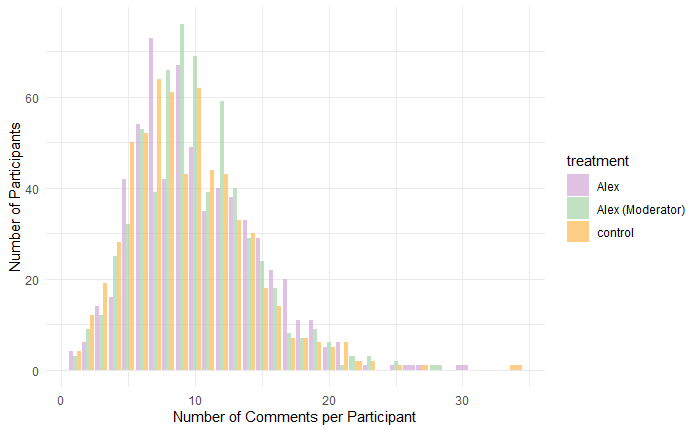}
    \caption{\emph{Distribution of the number of comments per participant across different treatments (Study 1).} The x-axis represents the number of comments made by each participant, while the y-axis shows the frequency of participants within each comment count. Each treatment is represented by a different color.}
    \label{fig:arguments_comments1}
\end{figure}

\newpage

\begin{figure}[ht]
    \centering
    \includegraphics[width=1\linewidth]{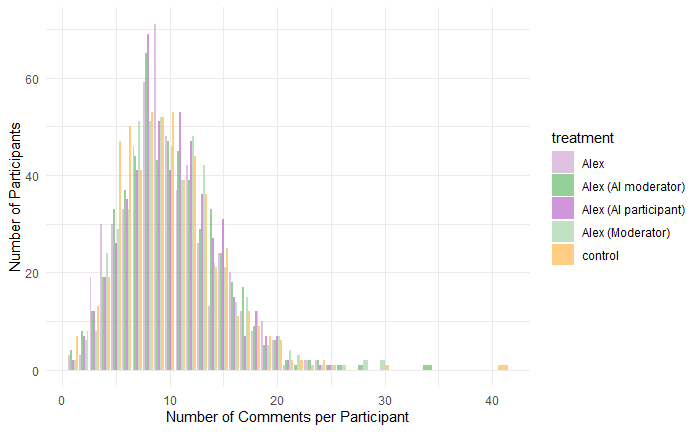}
    \caption{\emph{Distribution of the number of comments per participant across different treatments (Study 2).} The x-axis represents the number of comments made by each participant, while the y-axis shows the frequency of participants within each comment count. Each treatment is represented by a different color.}
    \label{fig:arguments_comments2}
\end{figure}

\newpage

\begin{figure} [ht]
    \centering
    \includegraphics[width=1\linewidth]{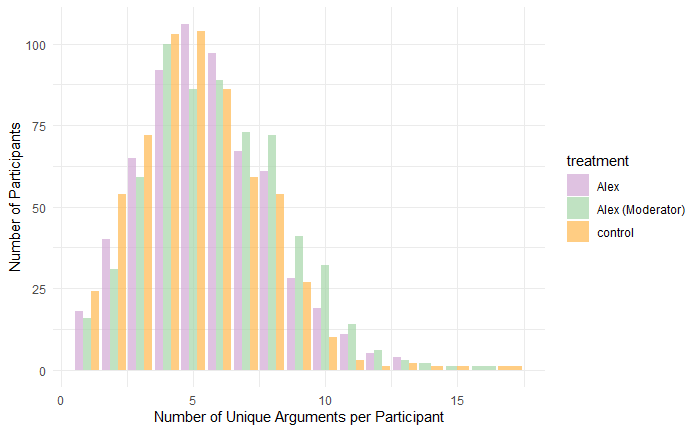}
    \caption{\emph{Distribution of the number of unique arguments per participant across different treatments (Study 1).} The x-axis represents the number of unique arguments made by each participant, while the y-axis shows the frequency of participants within each argument count. Each treatment is represented by a different color.}
    \label{fig:arguments_participant1}
\end{figure}

\newpage

\begin{figure}[ht]
    \centering
    \includegraphics[width=1\linewidth]{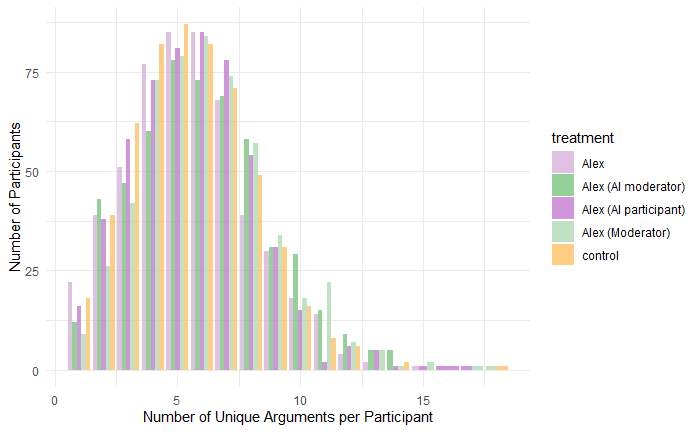}
    \caption{\emph{Distribution of the number of unique arguments per participant across different treatments (Study 2).} The x-axis represents the number of unique arguments made by each participant, while the y-axis shows the frequency of participants within each argument count. Each treatment is represented by a different color.}
    \label{fig:arguments_participant2}
\end{figure}

\newpage

\begin{figure}[ht]
    \centering
    \includegraphics[width=1\linewidth]{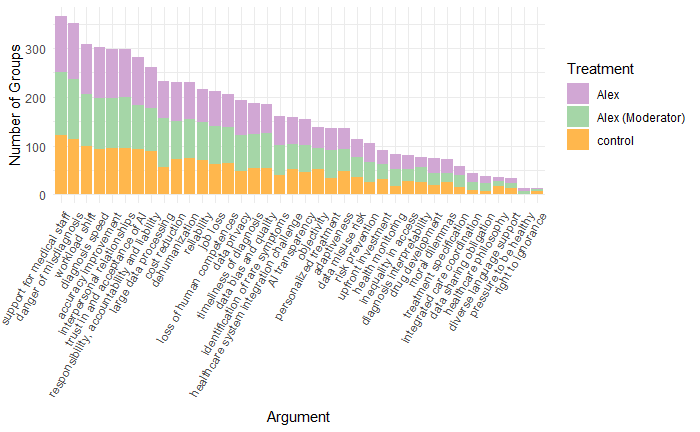}
    \caption{\emph{Number of groups mentioning each argument across different treatments (Study 1).} The x-axis represents the different arguments, and the y-axis shows the number of groups that mentioned each argument. Each treatment is represented by a different color.}
    \label{fig:arguments_distribution1}
\end{figure}

\newpage

\begin{figure}[ht]
    \centering
    \includegraphics[width=1\linewidth]{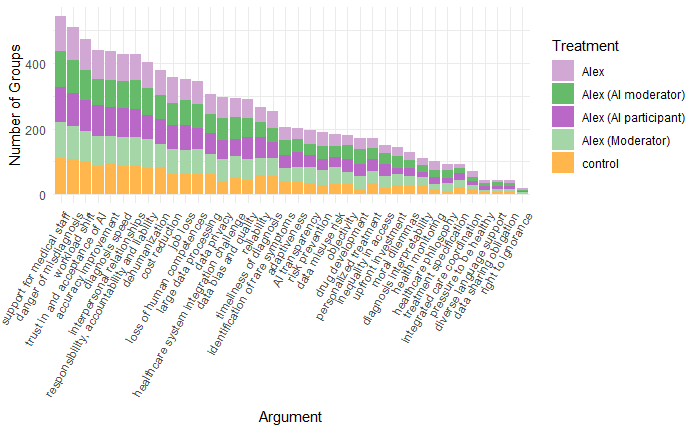}
    \caption{\emph{Number of groups mentioning each argument across different treatments (Study 2).} The x-axis represents the different arguments, and the y-axis shows the number of groups that mentioned each argument. Each treatment is represented by a different color.}
    \label{fig:arguments_distribution2}
\end{figure}


\newpage

\centering
\begin{longtable}{|p{2.5 cm}|p{3 cm}|p{7.5 cm}|}
    \caption{List of Arguments} \label{tab:listarguments} \\  
    \hline
    \textbf{Argument} & \textbf{Brief Explanation} & \textbf{Extensive Explanation} \\
    \hline\hline
    \endfirsthead
    \textit{large data processing} & AI can manage and process vast amounts of data more effectively. & AI is particularly adept at processing and organizing large amounts of medical data, such as patient records and lab results, which can be overwhelming for human professionals. This capability is critical because the sheer volume of information in healthcare is far beyond what humans can efficiently process.  AI serves as a powerful tool to support the healthcare system by enabling better management of large amounts of data. \\
    \hline
    \textit{diagnosis speed} & AI can speed up the diagnosis process in healthcare by quickly identifying illnesses. & AI can analyze medical images and test results much faster to detect signs of disease that would take humans much longer to detect. This speed is critical for making early and accurate diagnoses, so that treatment can begin promptly. Overall, AI's rapid analysis supports faster decision-making, which is critical for effective patient care. \\
    \hline
    \textit{identification of rare symptoms} & AI is good at spotting rare symptoms in healthcare data that humans might miss. & AI is capable of analyzing large data sets to identify rare or subtle signs of disease that are often missed by human clinicians. Its ability to process information from multiple sources, such as genetic data and patient histories, allows it to efficiently detect rare diseases.  \\
    \hline
    \textit{timeliness of diagnosis} & AI contributes to more timely diagnoses in healthcare and can thus help to detect diseases at an early stage. & With the help of AI, healthcare providers can achieve faster diagnosis times by automating the initial analysis of clinical data and test results. This reduces patient wait times and allows for earlier intervention, which can be critical in the management and treatment of various medical conditions. AI's rapid processing capabilities ensure that diagnoses are not only made quickly, but also based on the most recent data available, improving overall treatment effectiveness. \\
    \hline
    \textit{workload shift} & AI helps free up medical staff's time, allowing them to focus on patient care and professional development. & AI automates routine and data-intensive healthcare tasks, significantly reducing the workload of healthcare professionals. This saved time can be redirected to more direct patient care, in-depth medical research, or professional development and training. By handling the bulk of the data processing, AI enables healthcare workers to engage in more hands-on practice or enhance their skills and knowledge, ultimately leading to better patient outcomes and more advanced healthcare services. \\
    \hline
    \textit{support for medical staff} & AI can act as a support system for medical staff by assisting with various clinical tasks. & AI can support medical staff by assisting with diagnostic procedures, patient monitoring, and even suggesting potential treatment pathways. By continuously learning from new data, AI systems evolve and become more effective at assisting the medical team, contributing significantly to the overall quality of healthcare. \\
    \hline
    \textit{interpersonal relationships} & Care provided solely by AI can lead to interpersonal relationships being neglected. & AI can lead to less direct interaction between patients and medical staff. This reduction in face-to-face contact can affect the quality of the patient-provider relationship, which is often built on trust and personal understanding. As AI takes on more roles in diagnosing and managing patient care, there's a risk that the crucial human element that comforts and reassures patients during their treatment will diminish. \\
    \hline
    \textit{danger of misdiagnosis} & AI can increase the risk of misdiagnosis, necessitating human validation which might not always be feasible. & AI carries the risk of misdiagnosis due to limitations in the learning capabilities, issues related to training data or data interpretation. This risk requires thorough validation by human medical professionals, which can be challenging in environments with high patient volumes, limited staff, or complex AI models. Relying on AI without sufficient human oversight can lead to errors that are not caught in time, especially in complex or atypical cases where nuanced human judgment is essential. \\
    \hline
    \textit{dehumanization} & AI can contribute to the dehumanization of healthcare by prioritizing efficiency over patient-centered care. & The integration of AI into healthcare settings can sometimes shift the focus from patient-centered care to operational efficiency and cost reduction. This might lead to a healthcare environment where decisions are driven more by data and algorithms than by individual patient needs and preferences. Such a shift could make patients feel like they are being treated by a system rather than cared for by empathetic human beings, potentially compromising the quality of care. \\
    \hline
    \textit{data privacy} & AI can pose significant risks to patient data privacy. & The use of AI in healthcare requires access to vast amounts of personal health information, which raises significant privacy concerns. If not properly managed, there is a risk that sensitive patient data could be exposed or misused. The complexity and volume of data handled by AI systems also make them attractive targets for cyberattacks, further jeopardizing patient privacy and trust in the healthcare system. In addition, the most powerful AI solutions are provided by private companies and are cloud-based. Therefore, data must be shared with private companies, and cloud services are additionally vulnerable to cyberattacks. \\
    \hline
    \textit{personalized treatment} & AI can enhance personalized treatment by tailoring healthcare to individual patient needs. & AI algorithms analyze vast amounts of patient data, including genetic information, lifestyle factors and past health records, to develop personalized treatment plans. This capability enables treatments that are specifically tailored to an individual's unique health profile, potentially increasing their effectiveness. AI's ability to integrate and learn from multiple data sources continuously improves its recommendations, making personalized medicine more precise and effective over time. \\
    \hline
    \textit{treatment specification} & AI can specify treatments by closely analyzing disease patterns and patient responses. & AI's detailed analysis of disease manifestations and patient outcomes can help refine treatment protocols. By identifying the most effective interventions based on similar patient histories and characteristics, AI can suggest specific treatments that are likely to yield the best results. This level of specification not only increases treatment efficacy, but also minimizes the risk of adverse reactions, contributing to safer patient care. \\
    \hline
    \textit{health monitoring} & AI can improve health monitoring by continuously tracking patient health data. & AI tools in health monitoring devices collect and analyze real-time data such as heart rate, blood pressure, and other vital signs. This continuous monitoring helps in early detection of potential health issues, allowing for timely medical intervention. AI’s capability to alert both patients and healthcare providers about critical changes in health metrics ensures that necessary actions can be taken promptly, thus preventing complications. \\
    \hline
    \textit{accuracy improvement} & AI can improve the accuracy of diagnoses and the comprehensiveness of treatments in healthcare. & AI systems use advanced algorithms to analyze medical images, lab results, and other diagnostic tools with high accuracy. This reduces human error and can increase the reliability of diagnoses. In addition, AI's ability to learn from each case it processes helps to continually refine its diagnostic and treatment algorithms, improving the overall accuracy of medical services. \\
    \hline
    \textit{cost reduction} & AI can reduce costs in healthcare by streamlining operations and reducing resource utilization. & AI introduces efficiencies in healthcare management by automating routine tasks, optimizing treatment plans, and reducing the need for repetitive diagnostic tests. This not only speeds up the treatment process but also significantly cuts down on operational costs. Furthermore, AI can help in predicting patient admission rates and managing hospital resources more effectively, which in turn contributes to overall cost savings in healthcare infrastructure. \\
    \hline
    \textit{diagnosis interpretability} & AI can make it difficult to understand how diagnoses are reached. & AI systems, particularly those based on deep learning, often operate as "black boxes," where the decision-making process is not easily understood by humans. This lack of interpretability can be problematic, especially when healthcare providers need to explain diagnoses and treatment rationales to patients, potentially undermining trust and adherence to prescribed therapies. \\
    \hline
    \textit{data bias and quality} & AI can reflect issues in data quality, including bias, inaccuracy, and incompleteness, affecting healthcare decisions. & AI systems depend on the quality of the data they process, meaning any issues with data accuracy, currency, or completeness can be mirrored in the AI's output. Biases in training data, outdated information, or incomplete datasets can lead to flawed assumptions and decisions by the AI. These data quality issues not only compromise the reliability of AI-generated insights but can also lead to unequal healthcare outcomes across different patient groups, reinforcing existing disparities in healthcare access and treatment. \\
    \hline
    \textit{responsibility, accountability and liability} & Determining responsibility for AI decisions in healthcare can be complex. & When AI systems contribute to medical errors or adverse outcomes, it can be challenging to assign responsibility. This ambiguity affects accountability and liability, raising legal and ethical questions about who is at fault—the developers, the healthcare providers, or the AI itself—complicating the governance of AI use in healthcare. \\
    \hline
    \textit{reliability} & AI can offer more reliable performance in healthcare, unaffected by human limitations. & AI systems provide a consistent level of performance that is not influenced by factors that typically affect humans, such as fatigue, stress, or cognitive biases. This means AI can process data, recognize patterns, and suggest diagnostics with uniform precision at any time, under any workload. This reliability is especially valuable in healthcare settings where consistent accuracy is critical for patient care, ensuring that decisions and recommendations are based on data and algorithms rather than fluctuating human conditions. \\
    \hline
    \textit{healthcare system integration challenge} & AI integration into existing healthcare frameworks can be challenging. & Integrating AI technologies with current healthcare systems often involves overcoming substantial technical, cultural, and operational hurdles. These challenges include compatibility with legacy systems, training staff to use new AI tools, and modifying workflows, which can disrupt established practices and delay benefits. \\
    \hline
    \textit{trust in and acceptance of AI} & Building trust and acceptance of AI among healthcare professionals and patients can be difficult. & There is mistrust towards AI in healthcare among both patients and medical professionals, which can lead to low acceptance and integration of AI technologies. \\
    \hline
    \textit{upfront investment} & The initial investment for implementing AI in healthcare can be substantial. & Setting up AI systems in healthcare requires significant upfront costs, including expenditures for technology development or acquisition, system integration, and training personnel. These financial barriers can deter healthcare facilities, especially smaller ones, from adopting AI technologies. \\
    \hline
    \textit{loss of human competences} & Reliance on AI can lead to the erosion of traditional medical skills. & As AI takes over more routine and complex tasks, there's a risk that healthcare professionals may lose critical clinical skills. This dependency on technology could be detrimental in situations where AI assistance is unavailable or fails. \\
    \hline
    \textit{data misuse risk} & AI can be misused for patient profiling in ways that compromise ethics and privacy. & There is a risk that the detailed patient data handled by AI could be used for purposes beyond healthcare, such as marketing or insurance adjustments, without explicit patient consent. This misuse can violate privacy rights and ethical standards, leading to significant trust issues. \\
    \hline
    \textit{AI transparency} & Transparency in AI processes is often lacking in healthcare applications. & The opaque nature of many AI algorithms makes it hard for users and regulators to understand and trust the processes and outcomes. This lack of transparency can prevent stakeholders from fully assessing the fairness, effectiveness, and safety of AI applications in healthcare. \\
    \hline
    \textit{risk prevention} & AI can enhance risk prevention in healthcare by predicting and mitigating potential health issues. & AI algorithms are capable of analyzing large datasets to identify patterns that may indicate the risk of disease before symptoms become apparent. This proactive approach allows for earlier interventions that can prevent diseases from developing or worsening. Additionally, AI can monitor patient data in real-time, alerting healthcare providers to any changes that might require immediate action, thus enhancing patient safety and reducing the likelihood of medical emergencies. \\
    \hline
    \textit{drug development} & AI accelerates the drug development process, making it more efficient and less costly. & AI can significantly speed up the drug development cycle by simulating and predicting how new drugs will perform. This involves analyzing complex biochemical data and modeling drug interactions at a speed and scale unachievable by human researchers. By predicting the effectiveness and side effects of potential drugs earlier in the development process, AI reduces the time and resources needed to bring new treatments to market. \\
    \hline
    \textit{inequality in access} & AI can exacerbate healthcare access inequalities. & The deployment of AI in healthcare can widen the gap in access between different populations. While AI technologies have the potential to enhance healthcare efficiency and outcomes, they often require significant infrastructure and digital literacy, which are less accessible in under-resourced or rural areas. Additionally, the cost of implementing AI can be prohibitive, potentially leading to its primary use in wealthier urban centers or private practices. This disparity may result in uneven quality of care and differing health outcomes across socio-economic groups. \\
    \hline
    \textit{job loss} & AI can lead to job losses in the healthcare sector. & The integration of AI into healthcare settings can streamline many processes that were traditionally handled by human staff, such as data entry, diagnostic assistance, and even some aspects of patient interaction. While this can improve efficiency, it also poses a risk of reducing the need for certain roles, potentially leading to job losses. This impact might be particularly felt in administrative and some clinical support roles, altering employment patterns within the healthcare industry and necessitating a shift in workforce skills and training. \\
    \hline
    \textit{objectivity} & AI can offer objectivity in medical diagnoses and treatments, minimizing human biases. & AI systems make decisions based on data and predefined algorithms, largely unaffected by the personal biases or subjective interpretations that can influence human judgments. This objectivity ensures that treatment recommendations and diagnoses are consistent and based solely on the best available evidence, potentially leading to fairer and more standardized healthcare outcomes across diverse patient populations. \\
    \hline
    \textit{healthcare philosophy} & AI may prioritize quick, high-tech interventions over addressing underlying lifestyle factors in healthcare. & AI systems in healthcare are often optimized to suggest solutions that can be quantified and evaluated quickly, such as pharmaceutical treatments or the use of advanced medical equipment. This focus may lead to a preference for immediate fixes rather than exploring and supporting changes in lifestyle that could address the root causes of illnesses. Such an approach can sideline preventive medicine and holistic care strategies, potentially neglecting long-term health improvements in favor of short-term technological solutions. \\
    \hline
    \textit{adaptiveness} & AI may struggle to adapt to new medical scenarios or unexpected conditions. & AI can struggle when faced with novel or rare conditions not represented in its training set. This lack of adaptability can limit its usefulness in cases that deviate from "typical" scenarios, potentially leading to inappropriate or incorrect responses when flexibility and human intuition are critical. \\
    \hline
    \textit{diverse language support} & AI can facilitate healthcare delivery in multilingual contexts, improving accessibility. & In regions with diverse linguistic backgrounds, AI-powered tools can provide translation services and adapt communication to the patient's language, improving understanding and compliance. This capability ensures that language barriers do not impede access to quality healthcare, making information more accessible to patients and helping providers deliver more effective care. \\
    \hline
    \textit{pressure to be healthy} & AI-driven health monitoring can increase pressure on individuals to maintain constant health vigilance. & Continuous health monitoring by AI tools might create an environment where individuals feel pressured to constantly achieve optimal health metrics, potentially leading to anxiety and a sense of being surveilled. This pressure can detract from quality of life and may lead to unhealthy obsessions with personal health statistics instead of fostering a balanced approach to well-being. \\
    \hline
    \textit{data sharing obligation} & Widespread AI use in healthcare can pressure individuals into sharing their personal health data. & As AI becomes more integrated into healthcare systems, patients might feel compelled to share their personal health data to benefit from advanced AI-driven treatments and services. This perceived obligation can raise concerns, especially if individuals are uncomfortable with the extent of data sharing required. The pressure to share sensitive information can lead to discomfort and mistrust, particularly if the implications for privacy and data security are not clearly addressed or if the consent process is not robust and transparent. \\
    \hline
    \textit{moral dilemmas} & AI can introduce moral dilemmas in healthcare decisions, challenging ethical norms. & The use of AI in making healthcare decisions can lead to moral dilemmas, such as deciding between extending life at the cost of quality or determining eligibility for limited resources based on AI algorithms. These scenarios require nuanced ethical considerations that AI may not be equipped to handle, potentially leading to decisions that conflict with societal values and individual rights. \\
    \hline
    \textit{right to ignorance} & AI's capability to predict health outcomes can conflict with an individual’s right to not know certain information. & AI's predictive abilities can uncover information about potential future health conditions, which might be something an individual prefers not to know. Respecting this right to not know poses a challenge when using AI in healthcare, as it could inadvertently reveal sensitive information that impacts a person’s psychological well-being or future life choices. \\
    \hline
    \textit{integrated care coordination} & AI facilitates smoother and more effective information exchange among healthcare stakeholders. & AI systems can integrate and synthesize information from various sources, including medical records, research data, and real-time health monitoring, ensuring that all relevant stakeholders have access to comprehensive, up-to-date patient information. This improves collaboration and decision-making among different healthcare providers, payers, and researchers, leading to more coordinated care and better health outcomes. \\
    \hline
\end{longtable}

\newpage

\begin{table}[ht]
\centering
\footnotesize
\caption{\textbf{Results from Study 1 -- Number of unique arguments (objective)}. This table presents the results from linear regression models examining the relationships between various predictors and the "Number of unique Arguments" variable, with pooled results included for comparison.} \label{tab:study1_diversitymain}
\begin{tabular}{lcc}
\hline
Variable & Number of arguments & Number of arguments (Pooled) \\
\hline
Intercept & -0.212 & -0.136 \\
 & (0.383) & (0.383) \\
Pooled Effect &  & 0.244*** \\
 &  & (0.050) \\
Moderator & 0.332*** &  \\
 & (0.058) &  \\
Participant & 0.158** &  \\
 & (0.057) &  \\
Group size & -0.031 & -0.045 \\
 & (0.072) & (0.072) \\
Age & 0.001 & 0.001 \\
 & (0.002) & (0.002) \\
Sex (Male) & -0.181*** & -0.182*** \\
 & (0.047) & (0.047) \\
Education & 0.048* & 0.048** \\
 & (0.018) & (0.019) \\
Exp. political discussions & 0.032+ & 0.030 \\
 & (0.019) & (0.019) \\
Exp. online discussions & -0.029+ & -0.029+ \\
 & (0.017) & (0.017) \\
Num.Obs. & 1786 & 1786 \\
R2 & 0.034 & 0.029 \\
R2 Adj. & 0.029 & 0.025 \\
AIC & 5026.2 & 5033.6 \\
BIC & 5081.1 & 5083.0 \\
Log.Lik. & -2503.110 & -2507.804 \\
F & 7.747 & 7.481 \\
RMSE & 0.98 & 0.99 \\
\hline
\end{tabular}
\par \textit{Note: + \textit{p} \textless 0.1, * \textit{p} \textless 0.05, ** \textit{p} \textless 0.01, *** \textit{p} \textless 0.001}
\end{table}

\newpage 

\begin{table}[ht]
\centering
\footnotesize
\caption{\textbf{Results from Study 1 -- Range of viewpoints seen (subjective)}. This table presents the results from linear regression models examining the relationships between various predictors and the "Range of viewpoints" variable, with pooled results included for comparison.} \label{tab:study1_diversityviews}
\begin{tabular}{lcc}
\hline
Variable & Range of viewpoints & Range of viewpoints (Pooled) \\
\hline
Intercept & -0.821* & -0.833* \\
 & (0.383) & (0.382) \\
Pooled Effect &  & 0.037 \\
 &  & (0.050) \\
Moderator & 0.023 &  \\
 & (0.058) &  \\
Participant & 0.050 &  \\
 & (0.057) &  \\
Group size & 0.139+ & 0.141+ \\
 & (0.072) & (0.072) \\
Age & 0.002 & 0.002 \\
 & (0.002) & (0.002) \\
Sex (Male) & -0.074 & -0.074 \\
 & (0.047) & (0.047) \\
Education & -0.041* & -0.041* \\
 & (0.019) & (0.019) \\
Exp. political discussions & -0.034+ & -0.034+ \\
 & (0.019) & (0.019) \\
Exp. online discussions & 0.120*** & 0.120*** \\
 & (0.017) & (0.017) \\
Num.Obs. & 1786 & 1786 \\
R2 & 0.032 & 0.032 \\
R2 Adj. & 0.028 & 0.028 \\
AIC & 5029.1 & 5027.4 \\
BIC & 5084.0 & 5076.8 \\
Log.Lik. & -2504.565 & -2504.683 \\
F & 7.373 & 8.396 \\
RMSE & 0.98 & 0.98 \\
\hline
\end{tabular}
\par \textit{Note: + \textit{p} \textless 0.1, * \textit{p} \textless 0.05, ** \textit{p} \textless 0.01, *** \textit{p} \textless 0.001}
\end{table}

\newpage

\begin{table}[ht]
\centering
\footnotesize
\caption{\textbf{Results from Study 1 -- New arguments seen (subjective)}. This table presents the results from linear regression models examining the relationships between various predictors and the "New arguments" variables, with pooled results included for comparison.} \label{tab:study1_diversitynew}
\begin{tabular}{lcc}
\hline
Variable & New arguments & New arguments (Pooled) \\
\hline
Intercept & -0.813* & -0.787* \\
 & (0.379) & (0.378) \\
Pooled Effect &  & 0.206*** \\
 &  & (0.049) \\
Moderator & 0.236*** &  \\
 & (0.057) &  \\
Participant & 0.176** &  \\
 & (0.057) &  \\
Group size & 0.114 & 0.109 \\
 & (0.071) & (0.071) \\
Age & -0.007*** & -0.007*** \\
 & (0.002) & (0.002) \\
Sex (Male) & -0.073 & -0.073 \\
 & (0.046) & (0.046) \\
Education & 0.019 & 0.019 \\
 & (0.018) & (0.018) \\
Exp. political discussions & -0.013 & -0.014 \\
 & (0.018) & (0.018) \\
Exp. online discussions & 0.115*** & 0.115*** \\
 & (0.017) & (0.017) \\
Num.Obs. & 1786 & 1786 \\
R2 & 0.054 & 0.054 \\
R2 Adj. & 0.050 & 0.050 \\
AIC & 4987.8 & 4987.0 \\
BIC & 5042.7 & 5036.4 \\
Log.Lik. & -2483.923 & -2484.490 \\
F & 12.739 & 14.397 \\
RMSE & 0.97 & 0.97 \\
\hline
\end{tabular}
\par \textit{Note: + \textit{p} \textless 0.1, * \textit{p} \textless 0.05, ** \textit{p} \textless 0.01, *** \textit{p} \textless 0.001}
\end{table}

\newpage

\begin{table}[ht]
\centering
\footnotesize
\caption{\textbf{Results from Study 1 -- Participant's share of comments (objective).} This table presents the results from linear regression models examining the relationship between various predictors and the "Participant' share of comments" variable, with pooled results included for comparison.} \label{tab:study1_equalitymain}
\begin{tabular}{lcc}
\hline
Variable & Participant' share of comments & Participant' share of comments (Pooled) \\
\hline
(Intercept) & -0.648+ & -0.644+ \\
 & (0.385) & (0.384) \\
Pooled Effect &  & -0.004 \\
 &  & (0.050) \\
Moderator & 0.001 &  \\
 & (0.058) &  \\
Participant & -0.008 &  \\
 & (0.058) &  \\
Group size& 0.006 & 0.006 \\
 & (0.072) & (0.072) \\
age & 0.009*** & 0.009*** \\
 & (0.002) & (0.002) \\
Sex (Male) & -0.148** & -0.148** \\
 & (0.047) & (0.047) \\
Education & 0.008 & 0.008 \\
 & (0.019) & (0.019) \\
Exp. political discussions & 0.056** & 0.056** \\
 & (0.019) & (0.019) \\
Exp. online discussions & 0.024 & 0.024 \\
 & (0.017) & (0.017) \\
Num.Obs. & 1786 & 1786 \\
R2 & 0.026 & 0.026 \\
R2 Adj. & 0.022 & 0.022 \\
AIC & 5040.6 & 5038.6 \\
BIC & 5095.4 & 5088.0 \\
Log.Lik. & -2510.280 & -2510.291 \\
F & 5.909 & 6.753 \\
RMSE & 0.99 & 0.99 \\
\hline
\end{tabular}
\par \textit{Note: + \textit{p} \textless 0.1, * \textit{p} \textless 0.05, ** \textit{p} \textless 0.01, *** \textit{p} \textless 0.001}
\end{table}

\newpage

\begin{table}[ht]
\centering
\footnotesize
\caption{\textbf{Results from Study 1 -- Opportunity to participate (subjective).} This table presents the results from linear regression models examining the relationship between various predictors and the "Opportunity to participate" variable, with pooled results included for comparison.} \label{tab:study1_equalitybalance}
\begin{tabular}{lcc}
\hline
Variable & Opportunity to participate & Opportunity to participate (Pooled) \\
\hline
(Intercept) & -0.397 & -0.396 \\
 & (0.386) & (0.385) \\
Pooled Effect &  & -0.023 \\
 &  & (0.050) \\
Moderator & -0.021 &  \\
 & (0.058) &  \\
Participant & -0.025 &  \\
 & (0.058) &  \\
Group size & 0.017 & 0.016 \\
 & (0.073) & (0.073) \\
Age & 0.006** & 0.006** \\
 & (0.002) & (0.002) \\
Sex (Male) & -0.092+ & -0.092+ \\
 & (0.047) & (0.047) \\
Education & -0.015 & -0.015 \\
 & (0.019) & (0.019) \\
Exp. political discussions & -0.016 & -0.016 \\
 & (0.019) & (0.019) \\
Exp. online discussions & 0.086*** & 0.086*** \\
 & (0.017) & (0.017) \\
Num.Obs. & 1786 & 1786 \\
R2 & 0.018 & 0.018 \\
R2 Adj. & 0.013 & 0.014 \\
AIC & 5055.8 & 5053.8 \\
BIC & 5110.7 & 5103.2 \\
Log.Lik. & -2517.899 & -2517.901 \\
F & 3.972 & 4.541 \\
RMSE & 0.99 & 0.99 \\
\hline
\end{tabular}
\par \textit{Note: + \textit{p} \textless 0.1, * \textit{p} \textless 0.05, ** \textit{p} \textless 0.01, *** \textit{p} \textless 0.001}
\end{table}

\newpage 


\begin{table}[ht]
\centering
\footnotesize
\caption{\textbf{Results from Study 1 -- Different backgrounds (subjective).} This table presents the results from linear regression models examining the relationship between various predictors and the "Different backgrounds" variable, with pooled results included for comparison.} \label{tab:study1_equalitybackground}
\begin{tabular}{lcc}
\hline
Variable & Different backgrounds & Different backgrounds (Pooled) \\
\hline
(Intercept) & -0.688+ & -0.698+ \\
 & (0.379) & (0.378) \\
Pooled Effect &  & -0.062 \\
 &  & (0.049) \\
Moderator & -0.073 &  \\
 & (0.057) &  \\
Participant & -0.051 &  \\
 & (0.057) &  \\
Group size& 0.101 & 0.103 \\
 & (0.071) & (0.071) \\
Age & -0.002 & -0.002 \\
 & (0.002) & (0.002) \\
Sex (Male) & 0.005 & 0.006 \\
 & (0.046) & (0.046) \\
Education & -0.043* & -0.043* \\
 & (0.018) & (0.018) \\
Exp. political discussions & -0.005 & -0.004 \\
 & (0.018) & (0.018) \\
Exp. online discussions & 0.154*** & 0.154*** \\
 & (0.017) & (0.017) \\
Num.Obs. & 1786 & 1786 \\
R2 & 0.056 & 0.056 \\
R2 Adj. & 0.051 & 0.052 \\
AIC & 4985.2 & 4983.3 \\
BIC & 5040.1 & 5032.7 \\
Log.Lik. & -2482.601 & -2482.674 \\
F & 13.087 & 14.943 \\
RMSE & 0.97 & 0.97 \\
\hline
\end{tabular}
\par \textit{Note: + \textit{p} \textless 0.1, * \textit{p} \textless 0.05, ** \textit{p} \textless 0.01, *** \textit{p} \textless 0.001}
\end{table}

\newpage


\begin{table}[ht]
\centering
\footnotesize
\caption{ \textbf{Results from Study 1 -- Representativeness (subjective).} This table presents the results from linear regression models examining the relationship between various predictors and the "Representativeness" variable, with pooled results included for comparison.} \label{tab:study1_representativeness}
\begin{tabular}{lcc}
\hline
Variable & Representativeness & Representativeness (Pooled) \\
\hline
(Intercept) & -0.011 & -0.035 \\
 & (0.382) & (0.381) \\
Pooled Effect &  & -0.085+ \\
 &  & (0.050) \\
Moderator & -0.113+ &  \\
 & (0.057) &  \\
Participant & -0.058 &  \\
 & (0.057) &  \\
Group size & -0.037 & -0.033 \\
 & (0.072) & (0.072) \\
Age & -0.002 & -0.002 \\
 & (0.002) & (0.002) \\
Sex (Male) & -0.076 & -0.075 \\
 & (0.047) & (0.047) \\
Education & -0.025 & -0.025 \\
 & (0.018) & (0.018) \\
Exp. political discussions & 0.010 & 0.011 \\
 & (0.019) & (0.019) \\
Exp. online discussions & 0.126*** & 0.126*** \\
 & (0.017) & (0.017) \\
Num.Obs. & 1786 & 1786 \\
R2 & 0.040 & 0.040 \\
R2 Adj. & 0.036 & 0.036 \\
AIC & 5014.4 & 5013.3 \\
BIC & 5069.3 & 5062.7 \\
Log.Lik. & -2497.202 & -2497.671 \\
F & 9.273 & 10.465 \\
RMSE & 0.98 & 0.98 \\
\hline
\end{tabular}
\par \textit{Note: + \textit{p} \textless 0.1, * \textit{p} \textless 0.05, ** \textit{p} \textless 0.01, *** \textit{p} \textless 0.001}
\end{table}

\newpage


\begin{table}[ht]
\centering
\footnotesize
\caption{\textbf{Results from Study 2 -- Number of unique arguments (objective)}. This table presents the results from linear regression models examining the relationships between various predictors and the "Number of unique arguments" variable, with pooled results included for comparison.} \label{tab:study2_diversitymain}
\begin{tabular}{lcc}
\hline
Variable & Number of arguments & Number of arguments (Pooled) \\
\hline
(Intercept) & -0.153 & -0.134 \\
 & (0.307) & (0.307) \\
Pooled &  & 0.107* \\
 &  & (0.048) \\
Moderator & 0.231*** &  \\
 & (0.061) &  \\
AI Moderator & 0.152* &  \\
 & (0.061) &  \\
Participant & -0.011 &  \\
 & (0.061) &  \\
AI Participant & 0.060 &  \\
 & (0.061) &  \\
Group size & -0.052 & -0.057 \\
 & (0.058) & (0.058) \\
Age & 0.000 & 0.000 \\
 & (0.002) & (0.002) \\
Sex (Male) & -0.116** & -0.121** \\
 & (0.040) & (0.040) \\
Education & 0.058*** & 0.060*** \\
 & (0.015) & (0.015) \\
Exp. political discussions & 0.048** & 0.049** \\
 & (0.015) & (0.015) \\
Exp. online discussions & -0.029* & -0.030* \\
 & (0.015) & (0.015) \\
Num.Obs. & 2611 & 2611 \\
R2 & 0.024 & 0.018 \\
R2 Adj. & 0.020 & 0.015 \\
AIC & 7368.9 & 7380.6 \\
BIC & 7439.3 & 7433.4 \\
Log.Lik. & -3672.439 & -3681.289 \\
F & 6.433 & 6.627 \\
RMSE & 0.99 & 0.99 \\
\hline
\end{tabular}
\par \textit{Note: + \textit{p} \textless 0.1, * \textit{p} \textless 0.05, ** \textit{p} \textless 0.01, *** \textit{p} \textless 0.001}
\end{table}

\newpage


\begin{table}[ht]
\centering
\footnotesize
\caption{\textbf{Results from Study 2 -- Range of viewpoints seen (subjective)}. This table presents the results from linear regression models examining the relationships between various predictors and the "Range of viewpoints seen" variable, with pooled results included for comparison.} \label{tab:study2_diversityviews}
\begin{tabular}{lcc}
\hline
Variable & Range of viewpoints & Range of viewpoints (Pooled) \\
\hline
(Intercept) & -0.666* & -0.674* \\
 & (0.307) & (0.306) \\
Pooled &  & 0.026 \\
 &  & (0.048) \\
Moderator & -0.016 &  \\
 & (0.061) &  \\
AI Moderator & 0.037 &  \\
 & (0.061) &  \\
Participant & 0.040 &  \\
 & (0.061) &  \\
AI Participant & 0.041 &  \\
 & (0.061) &  \\
Group size & 0.108+ & 0.110+ \\
 & (0.058) & (0.057) \\
Age & -0.001 & -0.001 \\
 & (0.002) & (0.002) \\
Sex (Male) & -0.002 & -0.001 \\
 & (0.040) & (0.040) \\
Education & -0.019 & -0.019 \\
 & (0.015) & (0.015) \\
Exp. political discussions & -0.023 & -0.023 \\
 & (0.015) & (0.015) \\
Exp. online discussions & 0.101*** & 0.101*** \\
 & (0.015) & (0.015) \\
Num.Obs. & 2611 & 2611 \\
R2 & 0.022 & 0.022 \\
R2 Adj. & 0.019 & 0.019 \\
AIC & 7373.3 & 7368.5 \\
BIC & 7443.7 & 7421.3 \\
Log.Lik. & -3674.649 & -3675.254 \\
F & 5.983 & 8.380 \\
RMSE & 0.99 & 0.99 \\
\hline
\end{tabular}
\par \textit{Note: + \textit{p} \textless 0.1, * \textit{p} \textless 0.05, ** \textit{p} \textless 0.01, *** \textit{p} \textless 0.001}
\end{table}

\newpage


\begin{table}[ht]
\centering
\footnotesize
\caption{\textbf{Results from Study 2 -- New arguments seen (subjective)}. This table presents the results from linear regression models examining the relationships between various predictors and the "New arguments seen" variable, with pooled results included for comparison.} \label{tab:study2_diversitynew}
\begin{tabular}{lcc}
\hline
Variable & New arguments & New arguments (Pooled) \\
\hline
(Intercept) & -0.096 & -0.086 \\
 & (0.303) & (0.303) \\
Pooled &  & 0.205*** \\
 &  & (0.048) \\
Moderator & 0.164**  & \\
 & (0.061) &  \\
AI Moderator &  0.173** & \\
 & (0.060) &  \\
Participant & 0.293*** &  \\
 & (0.060) &  \\
AI Participant & 0.189** & \\
 & (0.060)  & \\
Group size & -0.004 & -0.006 \\
 & (0.057) & (0.057) \\
Age & -0.008*** & -0.008*** \\
 & (0.002) & (0.002) \\
Sex (Male) & -0.121** & -0.120** \\
 & (0.039) & (0.039) \\
Education & -0.018 & -0.018 \\
 & (0.015) & (0.015) \\
Exp. political discussions & 0.022 & 0.022 \\
 & (0.015) & (0.015) \\
Exp. online discussions & 0.094*** & 0.094*** \\
 & (0.014) & (0.014) \\
Num.Obs. & 2611 & 2611 \\
R2 & 0.046 & 0.044 \\
R2 Adj. & 0.042 & 0.041 \\
AIC & 7309.9 & 7309.7 \\
BIC & 7380.3 & 7362.5 \\
Log.Lik. & -3642.953 & -3645.859 \\
F & 12.519 & 17.039 \\
RMSE & 0.98 & 0.98 \\
\hline
\end{tabular}
\par \textit{Note: + \textit{p} \textless 0.1, * \textit{p} \textless 0.05, ** \textit{p} \textless 0.01, *** \textit{p} \textless 0.001}
\end{table}

\newpage

\begin{table}[ht]
\centering
\footnotesize
\caption{\textbf{Results from Study 2 -- Participant's share of comments (objective)}. This table presents the results from linear regression models examining the relationships between various predictors and the "Participant's share of comments" variable, with pooled results included for comparison.} \label{tab:study2_equalitymain}
\begin{tabular}{lcc}
\hline
Variable & Participant's share of comments & Participant's share of comments (Pooled) \\
\hline
(Intercept) & -0.367 & -0.371 \\
 & (0.308) & (0.308) \\
Pooled &  & -0.007 \\
 &  & (0.048) \\
Moderator & -0.015 &  \\
 & (0.062) &  \\
AI Moderator & -0.011 &  \\
 & (0.061) &  \\
Participant & -0.006 &  \\
 & (0.061) &  \\
AI Participant & 0.002 &  \\
 & (0.061) &  \\
Group size& 0.002 & 0.003 \\
 & (0.058) & (0.058) \\
Age & 0.005** & 0.005** \\
 & (0.002) & (0.002) \\
Sex (Male) & -0.118** & -0.118** \\
 & (0.040) & (0.040) \\
Education & 0.017 & 0.016 \\
 & (0.015) & (0.015) \\
Exp. political discussions & 0.051*** & 0.051*** \\
 & (0.016) & (0.015) \\
Exp. online discussions & -0.017 & -0.017 \\
 & (0.015) & (0.015) \\
Num.Obs. & 2611 & 2611 \\
R2 & 0.013 & 0.013 \\
R2 Adj. & 0.009 & 0.010 \\
AIC & 7398.1 & 7392.2 \\
BIC & 7468.6 & 7445.0 \\
Log.Lik. & -3687.073 & -3687.115 \\
F & 3.463 & 4.941 \\
RMSE & 0.99 & 0.99 \\
\hline
\end{tabular}
\par \textit{Note: + \textit{p} \textless 0.1, * \textit{p} \textless 0.05, ** \textit{p} \textless 0.01, *** \textit{p} \textless 0.001}
\end{table}

\newpage

\begin{table}[ht]
\centering
\footnotesize
\caption{\textbf{Results from Study 2 -- Opportunity to participate (subjective)}. This table presents the results from linear regression models examining the relationships between various predictors and the "Opportunity to participate" variable, with pooled results included for comparison.} \label{tab:study2_equalitybalance}
\begin{tabular}{lcc}
\hline
Variable & Opportunity to participate & Opportunity to participate (Pooled) \\
\hline
(Intercept) & 0.163 & 0.146 \\
 & (0.308) & (0.307) \\
Pooled &  & -0.158** \\
 &  & (0.048) \\
Moderator & -0.203*** & \\
 & (0.061) &  \\
AI Moderator & -0.163** &  \\
 & (0.061) &  \\
Participant & -0.144* & \\
 & (0.061) &  \\
AI Participant & -0.121* &  \\
 & (0.061) &  \\
Group size & 0.038 & 0.042 \\
 & (0.058) & (0.058) \\
Age & 0.002 & 0.002 \\
 & (0.002) & (0.002) \\
Sex (Male) & -0.062 & -0.060 \\
 & (0.040) & (0.040) \\
Education & -0.060*** & -0.061*** \\
 & (0.015) & (0.015) \\
Exp. political discussions & -0.033* & -0.033* \\
 & (0.015) & (0.015) \\
Exp. online discussions & 0.038** & 0.039** \\
 & (0.015) & (0.015) \\
Num.Obs. & 2611 & 2611 \\
R2 & 0.017 & 0.017 \\
R2 Adj. & 0.013 & 0.014 \\
AIC & 7388.0 & 7383.9 \\
BIC & 7458.4 & 7436.8 \\
Log.Lik. & -3682.013 & -3682.973 \\
F & 4.487 & 6.139 \\
RMSE & 0.99 & 0.99 \\
\hline
\end{tabular}
\par \textit{Note: + \textit{p} \textless 0.1, * \textit{p} \textless 0.05, ** \textit{p} \textless 0.01, *** \textit{p} \textless 0.001}
\end{table}

\newpage 


\begin{table}[ht]
\centering
\footnotesize
\caption{\textbf{Results from Study 2 -- Different backgrounds (subjective)}. This table presents the results from linear regression models examining the relationships between various predictors and the "Different backgrounds" variable, with pooled results included for comparison.} \label{tab:study2_equalitybackground}
\begin{tabular}{lcc}
\hline
Variable & Different backgrounds & Different backgrounds (Pooled) \\
\hline
(Intercept) & 0.111 & 0.096 \\
 & (0.303) & (0.302) \\
Pooled &  & -0.058 \\
 &  & (0.047) \\
Moderator & -0.080 & \\
 & (0.060) &  \\
AI Moderator & -0.091 & \\
 & (0.060) & \\
Participant & -0.041 &  \\
 & (0.060) &  \\
AI Participant & -0.021 & \\
 & (0.060) & \\
Group size & -0.022 & -0.018 \\
 & (0.057) & (0.057) \\
Age & -0.003* & -0.003* \\
 & (0.002) & (0.002) \\
Sex (Male) & -0.012 & -0.011 \\
 & (0.039) & (0.039) \\
Education & -0.060*** & -0.061*** \\
 & (0.015) & (0.015) \\
Exp. political discussions & -0.010 & -0.011 \\
 & (0.015) & (0.015) \\
Exp. online discussions & 0.143*** & 0.144*** \\
 & (0.014) & (0.014) \\
Num.Obs. & 2611 & 2611 \\
R2 & 0.050 & 0.049 \\
R2 Adj. & 0.046 & 0.047 \\
AIC & 7298.7 & 7294.5 \\
BIC & 7369.1 & 7347.3 \\
Log.Lik. & -3637.364 & -3638.237 \\
F & 13.689 & 19.316 \\
RMSE & 0.97 & 0.97 \\
\hline
\end{tabular}
\par \textit{Note: + \textit{p} \textless 0.1, * \textit{p} \textless 0.05, ** \textit{p} \textless 0.01, *** \textit{p} \textless 0.001}
\end{table}

\newpage


\begin{table}[ht]
\centering
\footnotesize
\caption{\textbf{Results from Study 2 -- Representativeness (subjective)}. This table presents the results from linear regression models examining the relationships between various predictors and the "Representativeness" variable, with pooled results included for comparison.} \label{tab:study2_representativeness}
\begin{tabular}{lcc}
\hline
Variable & Representativeness & Representativeness (Pooled) \\
\hline
(Intercept) & 0.420 & 0.422 \\
 & (0.304) & (0.304) \\
Pooled &  & -0.150** \\
 &  & (0.048) \\
Moderator & -0.185** &  \\
 & (0.061) &  \\
AI Moderator & -0.193** &  \\
 & (0.061) &  \\
Participant & -0.074 &  \\
 & (0.061) &  \\
AI Participant & -0.148* &  \\
 & (0.060) &  \\
Group size & -0.034 & -0.034 \\
 & (0.057) & (0.057) \\
Age & -0.005*** & -0.005*** \\
 & (0.002) & (0.002) \\
Sex (Male) & -0.095* & -0.094* \\
 & (0.039) & (0.039) \\
Education & -0.065*** & -0.065*** \\
 & (0.015) & (0.015) \\
Exp. political discussions & 0.028+ & 0.028+ \\
 & (0.015) & (0.015) \\
Exp. online discussions & 0.091*** & 0.091*** \\
 & (0.014) & (0.014) \\
Num.Obs. & 2611 & 2611 \\
R2 & 0.039 & 0.037 \\
R2 Adj. & 0.035 & 0.034 \\
AIC & 7329.6 & 7328.3 \\
BIC & 7400.0 & 7381.1 \\
Log.Lik. & -3652.790 & -3655.164 \\
F & 10.474 & 14.277 \\
RMSE & 0.98 & 0.98 \\
\hline
\end{tabular}
\par \textit{Note: + \textit{p} \textless 0.1, * \textit{p} \textless 0.05, ** \textit{p} \textless 0.01, *** \textit{p} \textless 0.001}
\end{table}

\newpage

\begin{table}
\centering
\footnotesize
\caption{ \textbf{Multilevel Regression Study 1 -- Part 1.}  This table presents the results from multilevel regression models examining the relationships between various predictors and three key variables: Number of unique arguments (objective), range of viewpoints seen (subjective), and new arguments seen (subjective). The coefficients for each predictor are shown, along with standard errors in parentheses.} \label{tab:study1_multileveldiversity}
\begin{tabular}{l c c c}
\hline
Variable & Number of unique arguments & Range of viewpoints & New arguments \\
\hline
(Intercept) & -0.216 & -0.822* & -0.810* \\
 & (0.389) & (0.387) & (0.388) \\
Moderator & 0.332*** & 0.023 & 0.236*** \\
 & (0.059) & (0.059) & (0.059) \\
Participant & 0.158** & 0.051 & 0.176** \\
 & (0.059) & (0.058) & (0.059) \\
Group Size & -0.031 & 0.139+ & 0.113 \\
 & (0.074) & (0.073) & (0.073) \\
Age & 0.001 & 0.002 & -0.007*** \\
 & (0.002) & (0.002) & (0.002) \\
Sex (Male) & -0.179*** & -0.072 & -0.070 \\
 & (0.047) & (0.047) & (0.046) \\
Education & 0.048** & -0.041* & 0.019 \\
 & (0.018) & (0.019) & (0.018) \\
Exp. Political Discussions & 0.032+ & -0.034+ & -0.013 \\
 & (0.019) & (0.019) & (0.018) \\
Exp. Online Discussions & -0.028 & 0.120*** & 0.115*** \\
 & (0.017) & (0.017) & (0.017) \\
SD (Intercept group\_nr) & 0.109 & 0.091 & 0.135 \\
SD (Observations) & 0.979 & 0.982 & 0.965 \\
Num.Obs. & 1786 & 1786 & 1786 \\
R2 Marg. & 0.033 & 0.032 & 0.054 \\
R2 Cond. & 0.045 & 0.040 & 0.072 \\
AIC & 5078.6 & 5081.8 & 5039.8 \\
BIC & 5139.0 & 5142.2 & 5100.1 \\
ICC & 0.0 & 0.0 & 0.0 \\
RMSE & 0.97 & 0.98 & 0.95 \\
\hline
\end{tabular}
 \par \textit{Note: + \textit{p} \textless 0.1, * \textit{p} \textless 0.05, ** \textit{p} \textless 0.01, *** \textit{p} \textless 0.001}
\end{table}

\newpage 

\begin{table}
\centering
\footnotesize
\caption{\textbf{Multilevel Regression Study 1 -- Part 2.} This table presents the results from multilevel regression models examining the relationships between various predictors and four key variables: Participant's share of comments (objective), opportunity to participate (subjective), different backgrounds (subjective) and representativeness (subjective). The coefficients for each predictor are shown, along with standard errors in parentheses.} \label{tab:study1_multilevelother}
\begin{tabular}{l c c c c}
\hline
Variable & Participant's share of comments & Opportunity to participate & Different backgrounds & Representativeness \\
\hline
(Intercept) & -0.648+ & -0.397 & -0.682+ & -0.010 \\
 & (0.385) & (0.386) & (0.392) & (0.384) \\
Moderator & 0.001 & -0.021 & -0.070 & -0.112+ \\
 & (0.058) & (0.058) & (0.060) & (0.058) \\
Participant & -0.008 & -0.025 & -0.050 & -0.058 \\
 & (0.058) & (0.058) & (0.060) & (0.058) \\
Group Size & 0.006 & 0.017 & 0.102 & -0.037 \\
 & (0.072) & (0.073) & (0.074) & (0.072) \\
Age & 0.009*** & 0.006** & -0.002 & -0.002 \\
 & (0.002) & (0.002) & (0.002) & (0.002) \\
Sex (Male) & -0.148** & -0.092+ & 0.005 & -0.075 \\
 & (0.047) & (0.047) & (0.046) & (0.047) \\
Education & 0.008 & -0.015 & -0.044* & -0.025 \\
 & (0.019) & (0.019) & (0.018) & (0.018) \\
Exp. Political Discussions & 0.056** & -0.016 & -0.004 & 0.010 \\
 & (0.019) & (0.019) & (0.018) & (0.019) \\
Exp. Online Discussions & 0.024 & 0.086*** & 0.153*** & 0.126*** \\
 & (0.017) & (0.017) & (0.017) & (0.017) \\
SD (Intercept group\_nr) & 0.000 & 0.000 & 0.160 & 0.059 \\
SD (Observations) & 0.989 & 0.993 & 0.961 & 0.980 \\
Num.Obs. & 1786 & 1786 & 1786 & 1786 \\
R2 Marg. & 0.026 & 0.017 & 0.055 & 0.040 \\
R2 Cond. & 0.045 & 0.040 & 0.072 & 0.043 \\
AIC & 5093.4 & 5108.6 & 5036.1 & 5067.3 \\
BIC & 5153.8 & 5168.9 & 5096.4 & 5127.7 \\
ICC & 0.0 & 0.0 & 0.0 & 0.0 \\
RMSE & 0.99 & 0.99 & 0.95 & 0.98 \\
\hline
\end{tabular}
\par \textit{Note: + \textit{p} \textless 0.1, * \textit{p} \textless 0.05, ** \textit{p} \textless 0.01, *** \textit{p} \textless 0.001}
\end{table}

\newpage

\begin{table}
\centering
\scriptsize
\caption{ \textbf{Multilevel Regression Study 2 -- Part 1.} This table presents the results from multilevel regression models examining the relationships between various predictors and three key variables: Number of unique arguments (objective), range of viewpoints seen (subjective), and new arguments seen (subjective). The coefficients for each predictor are shown, along with standard errors in parentheses.} \label{tab:study2_multileveldiversity}
\begin{tabular}{lcccc}
\hline
Variable & Number of unique arguments & Range of viewpoints & New arguments \\
\hline
(Intercept) & -0.166 & -0.669* & -0.093 \\
 & (0.319) & (0.329) & (0.319) \\
Moderator & 0.233*** & -0.014 & 0.164* \\
 & (0.065) & (0.068) & (0.065) \\
AI Moderator & 0.151* & 0.040 & 0.174** \\
 & (0.065) & (0.067) & (0.065) \\
Participant & -0.010 & 0.040 & 0.292*** \\
 & (0.065) & (0.067) & (0.065) \\
AI Participant & 0.059 & 0.043 & 0.190** \\
 & (0.065) & (0.067) & (0.065) \\
Group Size & -0.052 & 0.109+ & -0.004 \\
 & (0.060) & (0.062) & (0.060) \\
Age & 0.000 & -0.001 & -0.008*** \\
 & (0.002) & (0.002) & (0.002) \\
Sex (Male) & -0.116** & 0.006 & -0.120** \\
 & (0.039) & (0.039) & (0.039) \\
Education & 0.060*** & -0.020 & -0.019 \\
 & (0.015) & (0.015) & (0.015) \\
Exp. Political Discussions & 0.049** & -0.026+ & 0.021 \\
 & (0.015) & (0.015) & (0.015) \\
Exp. Online Discussions & -0.028+ & 0.102*** & 0.094*** \\
 & (0.015) & (0.015) & (0.014) \\
SD (Intercept group\_nr) & 0.179 & 0.238 & 0.197 \\
SD (Observations) & 0.973 & 0.962 & 0.959 \\
Num.Obs. & 2611 & 2611 & 2611 \\
R2 Marg. & 0.024 & 0.023 & 0.046 \\
R2 Cond. & 0.056 & 0.079 & 0.084 \\
AIC & 7427.7 & 7422.3 & 7366.5 \\
BIC & 7504.0 & 7498.6 & 7442.7 \\
ICC & 0.0 & 0.1 & 0.0 \\
RMSE & 0.96 & 0.94 & 0.94 \\
\hline
\end{tabular}
\par \textit{Note: + \textit{p} \textless 0.1, * \textit{p} \textless 0.05, ** \textit{p} \textless 0.01, *** \textit{p} \textless 0.001}
\end{table}

\newpage 

\begin{table}
\centering
\scriptsize
\caption{ \textbf{Multilevel Regression Study 2 -- Part 2.} This table presents the results from multilevel regression models examining the relationships between various predictors and four key variables: Participant's share of comments (objective), opportunity to participate (subjective), different backgrounds (subjective) and representativeness (subjective). The coefficients for each predictor are shown, along with standard errors in parentheses.} \label{tab:study2_multilevelother}
\begin{tabular}{lcccc}
\hline
Variable & Participant's share of comments & Opportunity to participate & Different backgrounds & Representativeness \\
\hline
(Intercept) & -0.367 & 0.163 & 0.118 & 0.418 \\
 & (0.308) & (0.313) & (0.316) & (0.306) \\
Moderator & -0.015 & -0.204** & -0.080 & -0.185** \\
 & (0.062) & (0.063) & (0.064) & (0.061) \\
AI Moderator & -0.011 & -0.163** & -0.089 & -0.193** \\
 & (0.061) & (0.063) & (0.064) & (0.061) \\
Participant & -0.006 & -0.144* & -0.041 & -0.074 \\
 & (0.061) & (0.063) & (0.064) & (0.061) \\
AI Participant & 0.002 & -0.121+ & -0.019 & -0.148* \\
 & (0.061) & (0.063) & (0.064) & (0.061) \\
Group Size & 0.002 & 0.038 & -0.022 & -0.034 \\
 & (0.058) & (0.059) & (0.060) & (0.057) \\
Age & 0.005*** & 0.002 & -0.003* & -0.005*** \\
 & (0.002) & (0.002) & (0.002) & (0.002) \\
Sex (Male) & -0.118** & -0.061 & -0.011 & -0.095* \\
 & (0.040) & (0.040) & (0.039) & (0.039) \\
Education & 0.017 & -0.061*** & -0.062*** & -0.065*** \\
 & (0.015) & (0.015) & (0.015) & (0.015) \\
Exp. Political Discussions & 0.051*** & -0.032* & -0.011 & 0.028+ \\
 & (0.016) & (0.015) & (0.015) & (0.015) \\
Exp. Online Discussions & -0.017 & 0.039** & 0.143*** & 0.091*** \\
 & (0.015) & (0.015) & (0.014) & (0.014) \\
SD (Intercept group\_nr) & 0.000 & 0.112 & 0.181 & 0.065 \\
SD (Observations) & 0.995 & 0.987 & 0.960 & 0.980 \\
Num.Obs. & 2611 & 2611 & 2611 & 2611 \\
R2 Marg. & 0.013 & 0.017 & 0.050 & 0.039 \\
R2 Cond. & 0.030 & 0.083 & 0.043 \\
AIC & 7461.7 & 7450.9 & 7357.4 & 7393.3 \\
BIC & 7538.0 & 7527.1 & 7433.7 & 7469.6 \\
ICC & 0.0 & 0.0 & 0.0 & 0.0 \\
RMSE & 0.99 & 0.98 & 0.94 & 0.98 \\
\hline
\end{tabular}
\par \textit{Note: + \textit{p} \textless 0.1, * \textit{p} \textless 0.05, ** \textit{p} \textless 0.01, *** \textit{p} \textless 0.001}
\end{table}

\newpage

\begin{table}
\centering
\footnotesize
\caption{ \textbf{Negative Binomial Regression for 'Number of unique arguments'.} This table presents the results of negative binomial regression models for 'Number of unique arguments', with separate analyses for Study 1 and Study 2. }\label{tab:negativebinomial}
\begin{tabular}{l c c}
\hline
Variable & Study 1 & Study 2 \\
\hline
(Intercept) & 1.607*** & 1.675*** \\
 & (0.173) & (0.141) \\
Moderator & 0.152*** & 0.105*** \\
 & (0.026) & (0.028) \\
AI moderator &  & 0.070* \\
 &  & (0.028) \\
Participant & 0.075** & -0.005 \\
 & (0.027) & (0.029) \\
AI participant &  & 0.029 \\
 &  & (0.028) \\
Group Size & -0.014 & -0.024 \\
 & (0.033) & (0.026) \\
Age & 0.000 & 0.000 \\
 & (0.001) & (0.001) \\
Sex (Male) & -0.083*** & -0.054** \\
 & (0.021) & (0.018) \\
Education & 0.022** & 0.027*** \\
 & (0.009) & (0.007) \\
Ep. Political Discsussions & 0.015+ & 0.022** \\
 & (0.009) & (0.007) \\
Exp. Online Discussions & -0.013+ & -0.013+ \\
 & (0.008) & (0.007) \\
Num.Obs. & 1786 & 2611 \\
AIC & 8310.0 & 12404.3 \\
BIC & 8364.8 & 12474.7 \\
Log.Lik. & -4144.984 & -6190.143 \\
F & 7.7550 & 6.387 \\
RMSE & 2.49 & 2.62 \\
\hline
\end{tabular}
\par \textit{Note: + \textit{p} \textless 0.1, * \textit{p} \textless 0.05, ** \textit{p} \textless 0.01, *** \textit{p} \textless 0.001}
\end{table}

\newpage

\begin{table}
\centering
\footnotesize
\caption{ \textbf{Model Results for 'Participant's share of comments' with Tokens.} This table presents the results of regression models for the 'Participant's share of comments' measure with tokens, comparing findings across Study 1 and Study 2.} \label{tab:equality_tokens}
\begin{tabular}{l c c}
\hline
 & Study 1 & Study 2 \\
\hline
Intercept & -0.405 & -0.264 \\
 & (0.388) & (0.308) \\
Moderator & 0.006 & -0.012 \\
 & (0.058) & (0.061) \\
AI moderator &  & -0.014 \\
 &  & (0.061) \\
Participant & 0.001 & -0.004 \\
 & (0.058) & (0.061) \\
AI participant &  & 0.007 \\
 &  & (0.061) \\
Group Size & 0.008 & -0.007 \\
 & (0.073) & (0.058) \\
Age & 0.002 & 0.000 \\
 & (0.002) & (0.002) \\
Sex (Male) & -0.106* & -0.079* \\
 & (0.047) & (0.040) \\
Education & 0.040* & 0.058*** \\
 & (0.019) & (0.015) \\
Exp. Political Discussions & 0.045* & 0.057*** \\
 & (0.019) & (0.016) \\
Exp. Online Discussions & -0.006 & -0.045** \\
 & (0.018) & (0.015) \\
Num.Obs. & 1786 & 2611 \\
R2 & 0.010 & 0.016 \\
R2 Adj. & 0.006 & 0.012 \\
AIC & 5068.8 & 7391.8 \\
BIC & 5123.7 & 7462.2 \\
Log.Lik. & -2524.404 & -3683.909 \\
F & 2.330 & 4.103 \\
RMSE & 0.99 & 0.99 \\
\hline
\end{tabular}
\par \textit{Note: * \textit{p} \textless 0.05, ** \textit{p} \textless 0.01, *** \textit{p} \textless 0.001}
\end{table}

\newpage

\begin{table}[ht]
\centering
\caption{\emph{Results of Contrasts Analysis.} This table presents the results of contrast analyses comparing various conditions in Study 1 and Study 2. For each contrast, the estimate, standard error (SE), degrees of freedom (df), t-ratio, p-value, and corresponding study are provided. } 

\begin{tabular}{lrrrrrl}
  \hline
contrast & estimate & SE & df & t.ratio & p.value & study \\ 
  \hline
Alex (Moderator) - Alex & 0.17 & 0.06 & 1777 & 3.06 & 0.00 & Study 1 \\ 
  Alex (Moderator) - Alex & 0.24 & 0.06 & 2596 & 3.92 & 0.00 & Study 2 \\ 
  Alex (AI participant) - Alex & 0.08 & 0.06 & 2596 & 1.29 & 0.57 & Study 2 \\ 
  Alex (AI participant) - Alex (Moderator) & -0.16 & 0.06 & 2596 & -2.61 & 0.04 & Study 2 \\ 
  Alex (AI moderator) - Alex & 0.16 & 0.06 & 2596 & 2.63 & 0.04 & Study 2 \\ 
  Alex (AI moderator) - Alex (Moderator) & -0.08 & 0.06 & 2596 & -1.28 & 0.57 & Study 2 \\ 
  Alex (AI moderator) - Alex (AI participant) & 0.08 & 0.06 & 2596 & 1.33 & 0.54 & Study 2 \\ 
   \hline
\end{tabular}
\label{tab:contrasts}
\end{table}

\begin{table}
    \centering
    \caption{\emph{Demographic Information of Participants.} This table provides the demographic characteristics of participants in Study 1 and Study 2. It includes the median age, age range, gender distribution, and Educational background for each study.}  
    \begin{tabular}{lccc}
    \hline
    \textbf{Demographic Variable} & \textbf{Category} & \textbf{Study 1 (n=1786)} & \textbf{Study 2 (n=2611)} \\ 
    \hline
    \textbf{Age} (Median) & - & 36 & 36 \\ 
    Age Range & - & 18 to 87 years & 18 to 82 years \\ 
    \textbf{Gender} & Male & 895 & 1175 \\ 
    (Absolute numbers) & Female & 891 & 1436 \\
    \textbf{Education} & without university degree & 558 & 822 \\ 
    (Absolute numbers) & with university degree & 1228 & 1789 \\ 
    \hline
    \end{tabular}
    \label{tab:demographics}
\end{table}

\end{document}